\documentclass[12pt,a4paper]{JHEP3} 

\usepackage{amsmath}
\usepackage{amssymb}
\usepackage{epsfig}
\usepackage{dcolumn}
\usepackage{enumerate}

\newcommand{\del}{\partial}

\title{QED in finite volume 
and finite size scaling effect on electromagnetic properties 
of hadrons} 

\author{Masashi Hayakawa\\ 
Department of Physics, Nagoya University, Nagoya, 
Aichi 464-8602, Japan\\ 
Email: \email{hayakawa@eken.phys.nagoya-u.ac.jp}
} 
 
\author{Shunpei Uno\\ 
Department of Physics, Nagoya University, Nagoya, 
Aichi 464-8602, Japan\\ 
Email: \email{uno@eken.phys.nagoya-u.ac.jp}}

\abstract{  
 On account of its application to the present and future 
analysis of the virtual photon correction 
to the hadronic properties by means of lattice QCD simulation, 
we initiate the study of the finite size scaling 
effect on the QED correction  
using low energy effective theory of QCD with QED. 
 For this purpose, we begin with formulating 
a new QED on the space with finite volume. 
 By adapting this formalism to the partially quenched QCD 
with electromagnetism, 
we explore the qualitative features 
of the finite size scaling effect on 
the electromagnetic correction 
to the masses of pseudo-Goldstone bosons. 
} 

\keywords{Chiral Lagrangians, 
Electromagnetic Processes and Properties, 
NLO Computations, 
Lattice QCD} 

\preprint 

\begin{document}

\section{Introduction} 
\label{sec:introduction}  

 The recent progress in the lattice QCD simulation 
enables us to perform ``measurements'' 
for hadronic properties 
through the world with more and more realistic QCD 
realized in the computer. 
 It may not be a far future that 
the lattice QCD simulation becomes a method 
by which precise measurement of 
nonperturbative dynamics of QCD is reached. 

 One of the important applications of the lattice QCD simulation 
would be the determination of quark masses. 
 The light quark masses have been determined 
using dynamical quarks with two flavors 
\cite{AliKhan:2001tx,Aoki:2002uc,Gockeler:2004rp, 
DellaMorte:2005kg,Becirevic:2005ta,Blossier:2007vv} 
and $(2 + 1)$ flavors  
\cite{Aubin:2004ck,Aubin:2004fs,
Mason:2005bj,Ishikawa:2007nn}. 
 The forthcoming precise measurement 
through the lattice QCD simulation 
reminds us that quarks are electrically charged. 
 All of the hadronic properties thus suffer from 
electromagnetic (EM) radiative corrections. 
 Because the EM interaction is other source 
of explicit breaking of isospin symmetry than 
the difference between 
the masses $m_u,\,m_d$ of up and down-quarks, 
the determination of $m_u - m_d$ requires us 
to grasp the size of EM correction quantitatively 
at the hadronic level. 
 Among the references 
\cite{AliKhan:2001tx,Aoki:2002uc,Gockeler:2004rp, 
DellaMorte:2005kg,Becirevic:2005ta,Blossier:2007vv, 
Aubin:2004ck,Aubin:2004fs,
Mason:2005bj,Ishikawa:2007nn}, 
Ref.~\cite{Aubin:2004fs} is the only work that presented 
the values of $m_u$, $m_d$ each, 
but it does not seem to make 
the data fit with a function parametrizing 
the EM splitting in the kaon masses as an undetermined constant. 

 The attempt to incorporate EM correction 
to the pseudoscalar meson masses  
in the context of lattice simulation was first done 
by Duncan {\it et al}.\cite{Duncan:1996xy} 
by simulating QED put on the lattice. 
 By adopting the same technique to incorporate QED correction, 
the QED correction to the meson masses 
has been investigated 
with use of dynamical domain wall fermions with two-flavors 
\cite{Blum:2007cy}, 
and in the quenched approximation 
with the renormalization group improved gauge action 
\cite{Namekawa:2005dr}. 
 Ref.~\cite{Shintani:2007ub} employs another method 
to calculate the leading-order EM correction to 
$\Delta m_\pi^2 \equiv m_{\pi^+}^2 - m_{\pi^0}^2$  
in the two-flavor overlap fermion simulation. 
 This work relies on the formula \cite{Das:1967it} 
\begin{eqnarray} 
 \left.\Delta m_\pi^2\right|_{\alpha}  
 &=& 
 \frac{3\,\alpha}{4\pi f_\pi^2} 
 \int \frac{d^4 k}{i \left(2\pi\right)} D^{\mu\nu}(k)\, 
  \nonumber \\ 
 && \quad 
 \times 
 \int d^4 x\,e^{i k \cdot x} 
 \left[ 
  \left<V^3_\mu(x)\,V^3_\nu(0)\right>_{\rm QCD} 
  - 
  \left<A^3_\mu(x)\,A^3_\nu(0)\right>_{\rm QCD} 
 \right] \, , 
  \label{eq:pionMassDiffSumRule}
\end{eqnarray} 
where $D^{\mu\nu}(k) = \eta^{\mu\nu} /(k^2 + i \epsilon)$ 
is the photon propagator,  
$f_\pi$ ($\simeq 92$ MeV) is the pion decay constant, 
$V^a_\mu \equiv \overline{q} T^a \gamma_\mu q$ 
and $A^a_\mu \equiv \overline{q} T^a \gamma_\nu \gamma_5 q$ 
with the normalization of $SU(2)$ generators 
${\rm tr}\,\left(T^a T^b\right) = \frac{1}{2}\,\delta^{ab}$. 
 $\left< \mathfrak{O} \right>_{\rm QCD}$ denotes 
the expectation value of the operator 
$\mathfrak{O}$ with respect to QCD. 
 The lattice simulation calculates 
the two correlation functions appearing on the right-hand side 
of Eq.~(\ref{eq:pionMassDiffSumRule}) 
and attempts to get the pion mass difference in the chiral limit 
\cite{Shintani:2007ub}. 

 Though our interest is the nonperturbative dynamics of QCD 
on space with infinite volume, 
the simulation has to be done in the virtual 
world with finite volume. 
 As quarks with colors and electric charges 
must live in such a finite volume, 
it is inevitable that QED correction 
suffers from finite size scaling effect 
no matter what computational method one may choose. 
 It is plausible that two pseudoscalar mesons 
in a common isospin multiplet share 
the same QCD finite size corrections 
\cite{Luscher:1983rk,Luscher:1985dn,Luscher:1986pf,Gasser:1987zq, 
Hasenfratz:1989pk,Hansen:1990un,Colangelo:2003hf, 
AliKhan:2003cu,Koma:2004wz,Colangelo:2004sc,Colangelo:2005gd, 
Caprini:2005zr,Colangelo:2006mp}. 
 In addition, electromagnetic force is long-ranged. 
 Therefore the finite size scaling effect regarding QED  
will dominate the finite size corrections in the EM splittings. 
 Unless it is adequately quantified, 
ignorance on the QED finite size scaling effect becomes 
another source of systematic uncertainty 
in $m_u - m_d$ derived using lattice simulation.  
 
 Thus far, there are diametrical opposite views 
on the relevance of QED finite size scaling. 
 The authors in Ref.~\cite{Namekawa:2005dr} performed 
the direct (quenched) lattice QCD measurement 
of the EM splitting in pion masses using 
two different sizes of four-volumes, 
$T \times L^3 = 24a \times (12a)^3$ and 
$24a \times (16a)^3$, where $a$ is the lattice spacing. 
 They observed no significant difference 
for the EM splitting measured in two volumes 
and concluded that 
a linear size $L = 2.4$ fm is sufficient to compute 
the EM splitting without correcting the measured values 
within available statistical uncertainty. 
 Contrastingly, Refs.~\cite{Duncan:1996xy,Blum:2007cy} 
estimated the QED finite size scaling according to 
the one-pole saturation approximation 
\cite{Das:1967it,Bardeen:1988zw,Ecker:1988te} 
to Eq.~(\ref{eq:pionMassDiffSumRule}) 
with the momentum integral replaced with the sum  
\begin{eqnarray} 
 \left.\Delta m_\pi^2\right|_{\alpha,\,{\rm VMD}}(T,\,L)  
 &=& 
 \frac{3 \alpha}{4\pi} \left(4\pi\right)^2 
 \frac{1}{T} 
 \frac{1}{L^3} 
 \sum_{k \in \left(\widetilde{\Gamma}_4 - \left\{0\right\}\right)} 
 \frac{m_V^2 m_A^2} 
      {k^2 
       \left(k^2 + m_V^2\right) 
       \left(k^2 + m_A^2\right)} \, , 
  \label{eq:naiveDiscretizationWithFiniteT} 
\end{eqnarray} 
where $m_V \simeq 770$ MeV, 
$m_A \simeq 970$ MeV is the mass of $A_1$ 
in the chiral limit, 
and 
\begin{eqnarray} 
 \widetilde{\Gamma}_4 
 &\equiv& 
 \left\{ 
  k = (k_0,\,k_1,\,k_2,\,k_3) 
  \left| 
   k_0 \in \frac{2\pi}{T}\,\mathbb{Z} \, ,  
   k_j \in \frac{2\pi}{L}\,\mathbb{Z}   
  \right.
 \right\} \, , 
\end{eqnarray} 
is a lattice on the Euclidean four-momentum space. 
 For the lattice geometry $T \times L^3 
= 32 a \times (16a)^3$ with $1/a \simeq 1.66$ GeV 
in Ref.~\cite{Blum:2007cy}, 
Eq.~(\ref{eq:naiveDiscretizationWithFiniteT}) leads 
\begin{eqnarray} 
 \frac{\left.\Delta m_\pi^2\right|_{\alpha,\,{\rm VMD}}(T,\,L)} 
      {\left.\Delta m_\pi^2\right|_{\alpha,\,{\rm VMD}} 
       (\infty,\,\infty)} 
 \simeq 0.9 \, , \label{eq:RBCvalue}
\end{eqnarray} 
which is not negligible. 

 Under such circumstances, 
we investigate the finite size scaling effect 
on the QED contribution to light pseudoscalar meson masses 
using the chiral perturbation theory 
including electromagnetism 
\cite{Urech:1994hd,Knecht:1997jw, 
Bijnens:2006mk,Haefeli:2007ey,Gasser:2003hk}. 
 Though the finial quantitative determination of 
the finite size correction must 
resort to the first principle calculation 
as in Ref.~\cite{Namekawa:2005dr}, 
any qualitative understanding as shown here will help one 
to make extrapolation to infinite volume  
in the future simulation study. 

 This paper actually consists of two parts. 
 In the first part, Sec.~\ref{sec:finiteVolumeQED}, 
we give a formulation of QED on the space with finite volume. 
 This task is necessary because a single classical charged particle 
does not satisfy the equation of motion, 
say, the Gauss' law constraint in the finite volume QED 
obtained via the ordinal compactification procedure. 
 This is intuitively understandable as 
the electric flux emanating from a nonzero charge 
finds nowhere else to go. 
 We thus begin with defining such a QED that accommodates 
a single charged particle on the space 
$\mathbb{R} \times \mathbb{T}^3$, 
where $\mathbb{T}^3$ is three-dimensional torus corresponding to 
the compact space. 
 The second part, Sec.~\ref{sec:FV-Meson},  
utilizes this QED 
to study the finite size scaling effect 
on the electromagnetic splittings in pseudoscalar meson masses. 
 For this purpose, 
the calculation is performed in the framework of 
the partially quenched chiral perturbation theory 
\cite{Bernard:1992mk,Sharpe:1992ft,Sharpe:2000bc,Sharpe:2001fh}
including the electromagnetism \cite{Bijnens:2006mk} 
in view of its practical application 
to the actual lattice simulation .  
 Sec.~\ref{sec:discussion} is denoted to 
discussion and conclusion. 
 Appendix \ref{sec:basicSums} collects 
the formulae for the basic sums which appear in the evaluation 
of finite size scaling effect. 

\section{QED in finite volume} 
\label{sec:finiteVolumeQED}
 The aim of this section is to present a new QED 
on the space with finite volume,  
which allows us to investigate 
the properties of a single charged particle. 
 We first clarify in Sec.~\ref{subsec:problem}
the problem itself that confronts us in the QED 
obtained by the ordinal compactification procedure. 
 In Sec.~\ref{subsec:newQED}, we define 
a new QED in finite volume and explain how it solves 
this problem. 
 Throughout this paper, 
the topology of the space is the three-dimensional torus 
$\mathbb{T}^3 \equiv \mathbb{S}^1 \times \mathbb{S}^1 
\times \mathbb{S}^1$ 
with a common circumference $L$ for all $\mathcal{S}^1$. 
 A point on $\mathbb{T}^3$ 
is thus specified 
by the coordinates ${\bf x} \equiv (x^1,\,x^2,\,x^3)$ 
obeying periodicity 
$x^j \cong x^j + L$ ($j = 1,\,2,\,3$).  
 As in the analysis of finite size scaling of QCD 
\cite{Luscher:1983rk,Luscher:1985dn,Luscher:1986pf}, 
the temporal direction $t = x^0$ is taken to be infinite, 
$t \in \mathbb{R}$, for single particle states 
to develop poles in the energy space. 
 We adopt the convention 
$\eta_{\mu\nu} = {\rm diag\left(1,\,-1,\,-1\,-1\right)}$ 
for the signature of the metric. 

\subsection{problem} 
\label{subsec:problem}
 The most familiar procedure to construct the corresponding theory 
on $\mathbb{T}^3$ is 
to impose periodic boundary conditions on all fields. 
 The electromagnetic theory 
obtained via such a na\"{i}ve $\mathbb{Z}^3$-orbifolding procedure 
\footnote{ 
 The word {\it na\"{i}ve} means here that 
the Fourier modes $\widetilde{A}_\mu(t,\,{\bf k} = {\bf 0})$ 
in Eq.~(\ref{eq:expansionOfGaugePotential}) 
are treated as basic variables. 
 We recall that the Wilson lines 
is appropriate variables. 
 The correct $\mathcal{Z}^3$-orbifolding 
would involve the integration over Wilson line 
$U_0(t) 
\sim 
\exp\left[ 
 i\,e \int dt\,\frac{1}{L^3}\,\widetilde{A}_0(t,\,{\bf 0}) 
\right]$ 
which leads 
to the constraint that the charges on every three-dimensional 
hypersurface should vanish in total  
\begin{eqnarray} 
 \frac{1}{L^3} \int d^3 {\bf x}\,j^0(t,\,{\bf x}) = 0 \, . 
\end{eqnarray} 
 We would like to see what happens in the 
na\"{i}ve procedure. 
}
is referred to as {\it QED$_{\mathbb{Z}^3}$} here. 
 In QED$_{\mathbb{Z}^3}$ the gauge potential $A_\mu(x)$ 
($x^\nu = (t,\,{\bf x})$), in particular, 
obeys periodic boundary condition in every spatial direction. 
 This form of boundary condition is motivated for 
the practical reason that the periodic or anti-periodic boundary 
condition is imposed along spatial directions 
for the available lattice QCD configurations 
and the conservation of quantized momenta  
at every QED vertex requires that every spatial component 
of photon momenta be an integer multiple of $\frac{2\pi}{L}$. 
 The action for the gauge kinetic term 
in QED$_{\mathbb{Z}^3}$ takes the usual form 
\begin{eqnarray} 
 S_\gamma &=& 
 \int dt \int_{\mathbb{T}^3} d^3{\bf x}\, 
 \left( 
  - \frac{1}{4}\,F_{\mu\nu}\,F^{\mu\nu} 
 \right) \, , \label{eq:actionPureEM}
\end{eqnarray} 
with the field strength 
$F_{\mu\nu} = \del_\mu A_\nu - \del_\nu A_\mu$. 

 To elucidate the problem from the practical point of view  
in our context, we go back to 
Eq.~(\ref{eq:naiveDiscretizationWithFiniteT}) 
for the EM splitting in the pion mass squared 
in the effective theory including 
the vector and axial-vector mesons.  
 The application of the $\mathbb{Z}^3$-orbifolding procedure 
to the effective Lagrangian including the vector 
and axial-vector mesons in Ref.~\cite{Ecker:1988te} 
immediately leads the same form of the Lagrangian 
written in terms of the spatially periodic fields. 
 This effective theory gives the expression for the EM splitting 
in Eq.~(\ref{eq:naiveDiscretizationWithFiniteT}) 
with $T \rightarrow \infty$ 
\begin{eqnarray} 
 \left.\Delta m_\pi^2\right|_{\alpha,\,{\rm VMD}}(\infty,\,L)  
 &=& 
 \frac{3 \alpha}{4\pi} \left(4\pi\right)^2 
 \int_{-\infty}^\infty \frac{dk^0}{2\pi}  
 \frac{1}{L^3} 
 \sum_{{\bf k} \in \widetilde{\Gamma}_3} 
 \frac{m_V^2 m_A^2} 
      {k^2 
       \left(k^2 + m_V^2\right) 
       \left(k^2 + m_A^2\right)} \, , 
  \label{eq:EMsplittingFiniteVolumeVMD} 
\end{eqnarray} 
where 
\begin{eqnarray} 
 \widetilde{\Gamma}_3 
 &\equiv& 
 \left\{ 
  {\bf k} = \left(k^1,\,k^2,\,k^3\right) 
  \left| 
   \,k^j \in \frac{2\pi}{L}\,\mathbb{Z} 
  \right.
 \right\} \, . \label{eq:spatialMometumLattice}
\end{eqnarray} 
 We can see that 
the quantity (\ref{eq:EMsplittingFiniteVolumeVMD}) 
has infrared (IR) divergence 
coming from the contribution of ${\bf k} = {\bf 0}$. 
 The expression 
(\ref{eq:EMsplittingFiniteVolumeVMD}) can be interpreted 
as the sum of the contribution of Kaluza-Klein modes 
in one-dimensional field theory. 
 The IR divergence appearing in Eq.~(\ref{eq:spatialMometumLattice}) 
is attributed to that of a massless mode from this point of view. 
 It should be reminded that 
$\left.\Delta m_\pi^2\right|_{\alpha,\,{\rm VMD}}$ in infinite 
volume is IR-finite. 
 Therefore, 
$\displaystyle{ 
\lim_{L \rightarrow \infty} 
\left.\Delta m_\pi^2\right|_{\alpha,\,{\rm VMD}}(\infty,\,L) 
\ne \left.\Delta m_\pi^2\right|_{\alpha,\,{\rm VMD}} 
}$. 
 The theory deducing Eq.~(\ref{eq:EMsplittingFiniteVolumeVMD}) 
is one of QED$_{\mathbb{Z}^3}$ 
with the content of matter fields 
and the action concretely specified. 
 The same pathology always emerges 
in every QED$_{\mathbb{Z}^3}$ theory 
irrespective of the details of the matter fields and the action. 
 Thus QED$_{\mathbb{Z}^3}$ cannot be used 
to study the finite size scaling effect on the EM splittings. 

 The origin of the above pathology will be traced back to 
the inconsistency of a single charged particle with 
the classical equation of motion. 
 Though it may be well-known, 
the detail observation of this point will help 
us to grasp the essence of our new QED. 
 Due to the periodicity of the gauge potential, 
the electromagnetic current 
$j^\mu(x) \equiv \delta S_{\rm matter} /\delta A_\mu(x)$ 
derived from the matter part $S_{\rm matter}$ of the action 
is also periodic along the spatial directions. 
 The current $j^\mu(x)$ is assumed to be conserved 
under the equations of motion derived 
from the variation of matter fields. 
 The classical equation of motion derived 
from the variation of $A_\mu(x)$ is hence 
\begin{eqnarray} 
 \del_\nu F^{\mu\nu}(x) = j^\mu(x) \, . 
\end{eqnarray} 
 This contains the Gauss' law constraint 
\begin{eqnarray} 
 {\bf \nabla} \cdot {\bf E}(x) = \rho(x) \, . 
  \label{eq:GaussLawConstr}
\end{eqnarray} 
where the electric field $E^j(x)$ 
and the charge density $\rho(x)$ are given by 
$E^j(x) = F^{0j}(x)$ and $\rho(x) = j^0(x)$, respectively. 
 For simplicity, 
we consider an infinitely heavy charged particle. 
 When it is at rest initially, the charge density 
and the profile of electric field are both constant in time 
\begin{eqnarray} 
 {\bf \nabla} \cdot {\bf E}({\bf x}) 
 = e\,\delta^3({\bf x}) \, . 
\end{eqnarray} 
 The inconsistency appears when both sides of this equation 
are integrated over the whole $\mathbb{T}^3$; 
the left-hand side vanishes while the right-hand side does not. 
 Likewise, we can see 
that any single charged particle cannot live on $\mathbb{T}^3$. 

\subsection{new QED on $\mathbb{R} \times \mathbb{T}^3$} 
\label{subsec:newQED}

 The observation in Sec.~\ref{subsec:problem} shows that 
the IR divergence in Eq.~(\ref{eq:EMsplittingFiniteVolumeVMD}) 
is a manifestation of inconsistency 
of a single charged particle with the classical equation of motion. 
 The aim of this section 
is to introduce an alternative QED in finite volume 
that solves this problem 
for the sake of our study of the finite size scaling effect 
on the EM splitting. 

 In the space $\mathbb{T}^3$, three-momenta 
take discrete values in $\widetilde{\Gamma}_3$ 
in Eq.~(\ref{eq:spatialMometumLattice}). 
 Accordingly, 
the gauge potential is decomposed in the Fourier series 
with respect to the spatial dimension  
\begin{eqnarray} 
 A_\mu(t,\,{\bf x}) &=& 
 \frac{1}{L^3} 
 \sum_{{\bf k} \in \widetilde{\Gamma}_3} 
 e^{i {\bf k} \cdot {\bf x}}\,\widetilde{A}_\mu(t,\,{\bf k}) 
  \, . \label{eq:expansionOfGaugePotential}
\end{eqnarray} 
 The new QED on $\mathbb{R} \times \mathbb{T}^3$, 
referred to as QED$_{L}$, is the theory 
without variables $\widetilde{A}_\mu(t,\,{\bf k} = {\bf 0})$ 
{\it ab initio}. 
 In other words, we do not incorporate 
the Wilson lines 
$U_\mu(t) \sim 
\exp\left[ 
 i\,e\,\int dt\,\frac{1}{L^3} \widetilde{A}_\mu(t,\,{\bf 0}) 
\right]$ as dynamical variables. 
 In terms of such a gauge potential, 
the action of pure electromagnetism is given by $S_\gamma$ 
in Eq.~(\ref{eq:actionPureEM}). 
 Below, we observe the various features possessed by QED$_{L}$. 

 First, we see how QED$_{L}$ solves 
the problems in Sec.~\ref{subsec:problem}. 
 The equation of motion is modified as follows. 
 Since the modes $\widetilde{A}_\mu(t,\,{\bf 0})$ 
are absent, 
the variation of the full action $S = S_\gamma + S_{\rm matter}$ 
with respect to the gauge potential becomes   
\begin{eqnarray} 
 \delta_A S &=& 
 \int_{-\infty}^\infty d t 
 \int_{\mathbb{T}^3} d^3 {\bf x}
 \left[ 
  - \frac{1}{2} \delta F_{\mu\nu} F^{\mu\nu}
  +  \delta A_\mu\,j^\mu 
 \right] \nonumber \\ 
 &=& 
 \int_{-\infty}^\infty d t 
 \sum_{{\bf k} \in \widetilde{\Gamma}_3^\prime} 
 \left[ 
  \delta A_\mu(t,\,{\bf k}) 
 \right. \nonumber \\ 
 && \qquad \qquad \qquad 
 \left. 
  \times 
  \int d^3 {\bf x}\,e^{i {\bf k} \cdot {\bf x}} 
  \left( 
    - \del_\nu F^{\mu\nu}(t,\,{\bf x}) + j^\mu(t,\,{\bf x}) 
  \right)
 \right] \, , 
\end{eqnarray} 
where 
\begin{eqnarray} 
 & 
 \displaystyle{ 
  \widetilde{\Gamma}_3^\prime 
  \equiv 
  \widetilde{\Gamma}_3 - \left\{{\bf 0}\right\} \, . 
 }& \label{eq:GammaTildePrime}
\end{eqnarray} 
 As the result, the extremum condition gives 
the equations of motion only for ${\bf k} \ne {\bf 0}$;  
\begin{eqnarray} 
 & 
 \displaystyle{ 
  \int d^3 {\bf x}\, 
  \cos\left({\bf k} \cdot {\bf x}\right) 
  \left\{ 
   \del_\nu F^{\mu\nu}(t,\,{\bf x}) 
    - j^\mu(t,\,{\bf x}) 
  \right\} = 0 \quad ({\bf k} \ne {\bf 0})\, . 
 }& \label{eq:equationOfMotionFromA}
\end{eqnarray} 
 The inconsistency seen in Sec.~\ref{subsec:problem} 
is therefore circumvented. 
 We also note 
that Eq.~(\ref{eq:equationOfMotionFromA}) 
instead of Eq.~(\ref{eq:GaussLawConstr}) 
no longer yields  
the equality between the charges contained in 
a domain $V$ in $\mathbb{T}^3$ 
and the electric flux penetrating the surface $\del V$. 

 The quantum field theory will be defined 
in the path integral framework by defining 
the measure of the gauge potential as usual. 
 Let $\delta \widetilde{A}_\mu(t,\,{\bf k})$ be an infinitesimal 
variation of the mode $\widetilde{A}_\mu(t,\,{\bf k})$ 
($t \in \mathbb{R}$,\, 
${\bf k} \in \widetilde{\Gamma}_3^\prime$). 
 The functional measure is defined corresponding to 
the norm in the space of gauge configurations 
in QED$_{L}$ 
\begin{eqnarray} 
 \left\|\delta A\right\|^2  
 \equiv 
 \int dt 
 \sum_{{\bf k} \in 
       \widetilde{\Gamma}_3^\prime} 
  \delta \widetilde{A}_\mu(t,\,{\bf k})\, 
  \delta \widetilde{A}^\mu(t,\,{\bf k}) \, . 
  \label{eq:normForGaugePotential}
\end{eqnarray} 
 As we will see later 
this norm will turn out to be gauge-invariant. 
 From the point of view of field theory on $\mathbb{R}$, 
$\widetilde{A}(t,\,{\bf 0})$ are massless fields. 
 Due to the absence of these modes in QED$_{L}$, 
the IR divergence as in Eq.~(\ref{eq:spatialMometumLattice}) 
no longer appears. 

 We next observe the gauge symmetry of QED$_{L}$. 
 The transformation of the gauge potential is given by 
\begin{eqnarray} 
 A_\mu(x) \mapsto 
 A^\prime_\mu(x) 
 = A_\mu(x) 
   + 
     \del_\mu \Lambda \, ,  
  \label{eq:gaugeTransformation}
\end{eqnarray} 
while a matter field $\Phi(x)$ with charge $Q_\Phi e$ 
is transformed as 
\begin{eqnarray} 
 \Phi(x) \mapsto \Phi^\prime(x) 
 &=& \exp[i Q_\Phi\,e\,\Lambda(x)]\,\Phi(x) \,  .
\end{eqnarray} 
 Here we assume that there is at least one matter field 
which has the minimum charge $e$. 
 The easiest way to identify the full gauge symmetry 
is to look at the full gauge symmetry of QED$_{\mathbb{Z}^3}$ 
first. 
 There, the general form of $\Lambda(x)$ 
that keeps the periodic boundary condition 
for the matter fields takes the form 
\begin{eqnarray} 
 \Lambda(x) &=& 
 \Lambda_P(x) 
 + 
 \frac{2\pi}{e\,L}\,\sum_{j=1}^3 m_j x^j 
 \quad 
 \left({\rm QED}_{\mathbb{Z}^3}\right) \, , 
\end{eqnarray} 
where $m_j \in \mathbb{Z}$, 
and $\Lambda_P(x)$ is periodic along $\mathbb{T}^3$.  

 Denoting 
the Fourier components of $\Lambda_P(t,\,{\bf x})$ by 
$\widetilde{\Lambda}_P(t,\,{\bf k})$, 
the gauge transformation for the gauge potential 
becomes in the three-momentum space  
\begin{eqnarray} 
 \widetilde{A}^\prime_j(t,\,{\bf k}) &=& 
 \widetilde{A}_j(t,\,{\bf k}) 
 + 
 \left( 
  i k^j \widehat{\Lambda}_P(t,\,{\bf k}) 
  + L^3\,\delta_{{\bf k},\,{\bf 0}}\, 
    \frac{2\pi}{e\,L}\,m_j 
 \right) 
 \quad ({\rm QED}_{\mathbb{Z}^3}) \, . 
\end{eqnarray}  
 In our new QED, $\widetilde{A}(t,\,{\bf 0})$ 
no longer exists. 
 It thus turns out that there are no redundancy 
corresponding to ${\bf m} \in \mathbb{Z}^3$ 
and time-dependent but spatially homogeneous 
part of $\Lambda_P(x)$, that is,  
\begin{eqnarray} 
 \widetilde{A}^\prime_j(t,\,{\bf k}) &=& 
 \widetilde{A}_j(t,\,{\bf k}) 
 + 
 i\,
 k^j \widehat{\Lambda}_P(t,\,{\bf k}) 
 \quad 
 \left({\rm QED}_{L}\right)\, , 
  \label{eq:gaugeTransformationNewQED}
\end{eqnarray}  
where  
\begin{eqnarray} 
 & 
 \displaystyle{ 
  \del_t \widetilde{\Lambda}_P(t,\,{\bf 0}) = 0 \, . 
 }& \label{eq:conditionOnGaugeParameter}
\end{eqnarray} 
 One can readily see 
that the set of all functions that fulfill 
Eq.~(\ref{eq:conditionOnGaugeParameter}) forms an abelian group. 
 From Eq.~(\ref{eq:gaugeTransformationNewQED}), 
it is also easy to see that 
this gauge group is exactly the redundancy 
that allows us to take the Coulomb gauge fixing condition 
\begin{eqnarray} 
 \del_j A_j(t,\,{\bf x}) = 0 \, . 
  \label{eq:CoulombGaugeCondition}
\end{eqnarray} 
 We recall that in QED$_{\mathbb{Z}^3}$  
a residual gauge symmetry survives even after 
imposing the condition (\ref{eq:CoulombGaugeCondition}), 
which should be fixed by the additional condition 
$\widetilde{A}_0(t,\,{\bf 0}) = 0$ \cite{Blum:2007cy}. 
 In contrast, 
the condition (\ref{eq:CoulombGaugeCondition}) 
suffices to fix redundancy in QED$_{L}$ 
leaving only the global symmetry. 
 Obviously, the norm (\ref{eq:normForGaugePotential}) 
in the space of gauge configurations is gauge-invariant. 
 We can also make a BRST complex by introducing the ghost fields 
corresponding to the gauge parameters of the form 
(\ref{eq:conditionOnGaugeParameter}) 
in the standard manner \cite{Henneaux:1992ig}. 

 We close this section with a few remarks. 
 We consider 
the scattering of a charged particle and its anti-particle 
in the center of mass frame, 
leaving aside the issue 
whether the scattering may not be well-defined 
in the presence of a long-ranged force in finite volume.  
 Due to the absence of the modes 
$\widetilde{A}_\mu(t,\,{\bf 0})$, 
the $s$-channel process 
mediated by a single virtual photon does not occur. 
 However, we recall that Lorentz invariance, 
in particular, the symmetry related to the Lorentz boosts, 
is explicitly violated on the space $\mathbb{R} \times \mathbb{T}^3$. 
 Thus, once we consider the collision 
say,  of an incident charged particle 
with three-momentum 
$(p + \frac{2\pi}{L}) {\bf e}_x$, 
where $p \in \frac{2\pi}{L} \mathbb{Z}$ and ${\bf e}_x$ 
is a unit three-vector along $x$-direction, 
and an anti-particle with momentum $(-p) {\bf e}_x$,   
the $s$-channel process occurs. 
 In the limit $L \rightarrow \infty$, 
the cross section will approach to that 
in the center of mass frame in infinite volume. 

 Secondly, 
since the equation of motion 
(\ref{eq:equationOfMotionFromA}) is not written locally, 
QED$_{L}$ seems to possess somewhat non-locality. 
 In fact, it is not still clear at the present stage 
whether this is actually the case, 
and then whether another pathology appears in QED$_{L}$. 
 As will be demonstrated explicitly in the subsequent section, 
even if non-locality is present, 
it is so mild that 
the structure of ultra-violet (UV) divergence in QED$_{L}$ 
remains completely the same as in QED in infinite volume.  
 This feature is contrasted 
the situation in noncommutative field theory 
\cite{Filk:1996dm,Minwalla:1999px,Hayakawa:1999yt}; 
the the UV structure of a noncommutative field theory 
differs significantly from that of the commutative counterpart 
due to the hard non-locality. 

\section{Finite size scaling in meson mass} 
\label{sec:FV-Meson} 
 Now we apply QED$_{L}$ to the study of 
finite size effect on the EM correction 
to the pseudoscalar meson masses 
in the chiral perturbation theory including 
the electromagnetism 
\cite{Urech:1994hd,Knecht:1997jw,Bijnens:2006mk,Haefeli:2007ey}. 
 Looking at the practical application to the lattice simulation, 
we adopt the partially quenched chiral perturbation theory 
\cite{Bernard:1992mk,Sharpe:1992ft,Sharpe:2000bc,Sharpe:2001fh}
including the electromagnetism 
\cite{Bijnens:2006mk} and compute 
the leading-order finite size correction to the EM splitting 
in this theory. 
 For that purpose, we begin with summarizing our notations 
for partially quenched chiral perturbation theory 
including electromagnetism to the next-leading order. 
 We derive the formulas for the next-to-leading 
order correction to the pseudoscalar meson mass 
in finite volume, 
and evaluate them numerically to investigate 
the finite size correction to the EM splitting. 

\subsection{partially quenched 
chiral perturbation theory with electromagnetism} 
 The next-to-leading order corrections 
to the off-diagonal meson masses has already been 
computed in Ref.~\cite{Bijnens:2006mk}. 
 However, the expression written by the momentum integrals 
are needed in practice 
for the study of finite size scaling effect. 
 To derive such an expression 
in Sec.~\ref{subsec:EMfiniteSizeCorrection}, 
we fix the notations, in particular, 
of the low-energy constants at the next-to-leading order 
for the subsequent calculation, 
and dictate the free meson propagators 
necessary for the one-loop calculation. 
 To take the application to the partially quenched system 
with two-flavors, 
the super-trace of the EM charge matrix 
$\widetilde{\mathcal{Q}}$ is not assumed here  
to vanish, unlike in Ref.~\cite{Bijnens:2006mk}. 
 There then appear more local terms at the next-to-leading order 
(i,e., $O(p^4)$, $O(e^2 p^2)$ and $O(e^4)$) 
than those found in Ref.~\cite{Bijnens:2006mk}. 
 We list up all of them 
by drawing upon Appendix of Ref.~\cite{Knecht:1997jw}  
where one-loop UV divergences 
were computed for generic flavor number $N_F$ 
in the unquenched chiral perturbation theory 
including electromagnetism. 
 After that, we compute the UV divergences 
that should be absorbed by the coefficients 
of these local terms. 

 In what follows, 
all the fields and parameters are written 
in the ``flavor'' basis  
\begin{eqnarray} 
 \mathfrak{Q} &\equiv& 
 \left( 
  q^V_1,\,\cdots,\,q^V_{N_V},\, 
  q^S_1,\,\cdots,\,q^S_{N_S},\, 
  g_1,\,\cdots,\,g_{N_V} 
 \right)^T \, , 
\end{eqnarray} 
where $q^S_r$ ($r=1,\,\cdots,\,N_S$) denote 
the sea quark fields, 
$q^V_\alpha$ ($\alpha=1,\,\cdots,\,N_V$) the valence quark fields, 
and $g_\alpha$ ($\alpha=1,\,\cdots,\,N_V$) the ghost quark fields 
\cite{Bernard:1992mk}. 
 The chiral symmetry is a graded Lie group 
$G = SU(N_S + N_V | N_V)_L \times SU(N_S + N_V | N_V)_R$. 
 It breaks down spontaneously to 
its vector-like subgroup $H = SU(N_S + N_V | N_V)_V$. 
 The associated Nambu-Goldstone bosons are represented 
by an $(N_S + N_V | N_V) \times (N_S + N_V | N_V)$ 
supermatrix $\Pi$. 
 Using 
\begin{eqnarray} 
 u[\Pi(x)] &=& 
 \exp\left(i\,\frac{\Pi(x)}{\sqrt{2}\,F_0}\right) \, , 
\end{eqnarray} 
$\Pi$ transforms nonlinearly under $(g_L,\,g_R) \in G$ through 
\begin{eqnarray} 
 u[\Pi] \mapsto u[\Pi^\prime] 
 = g_R\,u[\Pi]\,h((g_L,\,g_R);\,\Pi)^\dagger  
 = h((g_L,\,g_R);\,\Pi)\,u[\Pi]\,g_L^\dagger \, . 
\end{eqnarray} 
where $h((g_L,\,g_R);\,\Pi) \in H$. 
 We follow the convention for the chiral Lagrangian 
which can be read off 
from Refs.~\cite{Bijnens:2006mk,Bijnens:2006zp} 
with minor modification. 

 The external fields $R_\mu(x)$, $L_\mu(x)$
that couple to the right-handed and left-handed 
chiral components of vector currents 
are incorporated in the partially quenched QCD action 
for the purpose of calculating 
the connected Green functions 
of vector and axial-vector currents. 
 They are defined to transform 
under local $(g_L,\,g_R)$ as  
\begin{eqnarray} 
 & 
 \displaystyle{ 
  L_\mu \mapsto L^\prime_\mu 
   = g_L L_\mu\,g_L^\dagger + i g_L\,\del_\mu g_L^\dagger\, , 
 }& \nonumber \\ 
 & 
 \displaystyle{ 
  R_\mu \mapsto R^\prime_\mu 
  = g_R R_\mu\,g_R^\dagger + i g_R\,\del_\mu g_R^\dagger\, . 
 }& 
\end{eqnarray} 
 The field strengths 
\begin{eqnarray} 
 & 
 \displaystyle{ 
  L_{\mu\nu} = \del_\mu L_\nu - \del_\nu L_\mu - [L_\mu,\,L_\nu] 
  \, , \quad 
  R_{\mu\nu} = \del_\mu R_\nu - \del_\nu R_\mu - [R_\mu,\,R_\nu] 
  \, , 
 }& 
\end{eqnarray} 
hence transform covariantly. 
 The symmetry breaking parameters are promoted 
to the spurion fields with the appropriate transformation laws. 
 For instance, let $\mathcal{M}(x)$ be 
the spurion field corresponding to the quark mass matrix $M$ 
and generalize the mass term 
in the partially quenched QCD to the form   
\begin{eqnarray} 
 & 
 \displaystyle{ 
  - \overline{\mathfrak{Q}}_R\,\mathcal{M}\,\mathfrak{Q}_L 
  - \overline{\mathfrak{Q}}_L\,\mathcal{M}^\dagger\, 
   \mathfrak{Q}_R \, . 
 }& 
\end{eqnarray} 
 The spurion field is assumed to transform under 
the local $(g_L,\,g_R) \in G$ as  
\begin{eqnarray} 
 \mathcal{M} \mapsto 
 \mathcal{M}^\prime = g_R\,\mathcal{M}\,g_L^\dagger \, . 
\end{eqnarray} 
 We write up the low-energy effective Lagrangian 
in terms of Nambu-Goldstone boson fields 
whose generating functional of connected Green functions
exhibits the same transformation 
as that of the microscopic theory 
\cite{Gasser:1983yg,Gasser:1984gg}. 
 The parameters are then inserted 
at the positions compatible with 
the way how the chiral symmetry is broken by them 
in the Feynman diagrams in the low energy effective theory. 
 In our context the $U(1)_{\rm em}$-charge matrix $\mathcal{Q}$ 
must also be promoted to a pair of spurion fields 
$\mathcal{Q}_L$, $\mathcal{Q}_R$ which transform 
respectively as 
\begin{eqnarray} 
 & 
 \displaystyle{ 
  \mathcal{Q}_L \mapsto 
   \mathcal{Q}_L^\prime = g_L\,\mathcal{Q}_L\,g_L^\dagger \, , 
  \quad 
  \mathcal{Q}_R \mapsto 
   \mathcal{Q}_R^\prime = g_R\,\mathcal{Q}_R\,g_R^\dagger \, . 
 }& 
\end{eqnarray} 
 The trace of any integral multiple 
of $\mathcal{Q}_L$ ($\mathcal{Q}_R$) is chirally invariant. 
 It is thus possible to impose the chirally invariant condition 
\begin{eqnarray} 
 & 
 \displaystyle{ 
  {\rm str}\left(\mathcal{Q}_R\right) 
  = 
  {\rm str}\left(\mathcal{Q}_L\right) \, . 
 }& \label{eq:ConditionOnLRChargeMatrix} 
\end{eqnarray} 
 They are not required here to vanish. 
 After writing up the Lagrangian to the order of our interest, 
$\mathcal{M}$ is set to the diagonal quark mass matrix 
\begin{eqnarray} 
 M_d = 
 {\rm diag} 
 \left( 
  m^V_1,\,\cdots,\,m^V_{N_V},\, 
  m^S_1,\,\cdots,\,m^S_{N_S},\, 
  m^V_1,\,\cdots,\,m^V_{N_V} 
 \right) \, , \label{eq:diagonalQuarkMassMatrix}
\end{eqnarray} 
and $\mathcal{Q}_L$, $\mathcal{Q}_R$ 
are set to the diagonal EM charge matrix 
\begin{eqnarray} 
 \mathcal{Q} = 
 {\rm diag} \left( 
  Q^V_1,\,\cdots,\,Q^V_{N_V},\, 
  Q^S_1,\,\cdots,\,Q^S_{N_S},\, 
  Q^V_1,\,\cdots,\,Q^V_{N_V} 
 \right) \, . \label{eq:DiagonalchargeMatrix}  
\end{eqnarray} 
 The substitution 
\begin{eqnarray} 
 & 
 \displaystyle{ 
  L_\mu \mapsto L_\mu + e \mathcal{Q} A_\mu \, , 
  \quad 
  R_\mu \mapsto R_\mu + e \mathcal{Q} A_\mu \, . 
 }&  
\end{eqnarray} 
introduces the coupling of photons to the meson fields. 

 In practice, 
in order to write down chirally invariant operators,  
it is convenient to use 
the building blocks $\mathcal{O}$ which transform 
as $\mathcal{O} \mapsto 
h((g_L,\,g_R);\,\Pi)\,\mathcal{O}\,h((g_L,\,g_R);\,\Pi)^\dagger$. 
 A set of the building blocks $\mathcal{O}$, 
each of which is also the eigenstates of charge conjugation 
and intrinsic parity transformation,   
is 
\begin{eqnarray} 
 && 
 u_\mu 
 \equiv 
 i \left\{  
    u^\dagger \left(\del_\mu u - i R_\mu u\right) 
    - u \left(\del_\mu u^\dagger - i L_\mu u^\dagger\right) 
   \right\} \, , \nonumber \\ 
 && 
 \chi_\pm 
 \equiv u^\dagger \chi u^\dagger \pm u \chi^\dagger\,u \, , \quad 
 \chi \equiv 2 B_0\,\mathcal{M} \, , 
  \nonumber \\ 
 && 
 \widetilde{\mathcal{Q}}_L \equiv u \mathcal{Q}_L\,u^\dagger \, , 
 \quad 
 \widetilde{\mathcal{Q}}_R \equiv 
  u^\dagger\,\mathcal{Q}_R\,u \, , \nonumber \\ 
 && 
 \mathfrak{F}_{\pm\,\mu\nu} \equiv 
 u\,L_{\mu\nu}\,u^\dagger \pm u^\dagger\,R_{\mu\nu}\,u \, ,  
\end{eqnarray} 
where $B_0$ is the mass scale characterizing 
the size of the chiral condensate \cite{Gasser:1984gg}, 
and their covariant derivatives with respect to 
the Maurer-Cartan form 
\begin{eqnarray} 
 \nabla_\mu \mathcal{O} 
 &\equiv& 
 \del_\mu \mathcal{O} - i [\Gamma_\mu,\,\mathcal{O}] \, , 
  \nonumber \\ 
 \Gamma_\mu 
 &\equiv& 
 - \frac{1}{2} 
   \left\{ 
    u^\dagger \left(\del_\mu u - i R_\mu u\right) 
    + 
    u \left(\del_\mu u^\dagger - i L_\mu u^\dagger\right)
   \right\} \, . 
\end{eqnarray} 

 The chiral Lagrangian at the leading-order,  
$O(p^2) \sim O(e^2)$, takes the similar form 
as in the unquenched case 
\begin{eqnarray} 
 \mathcal{L}_2 &=& 
 \frac{F_0^2}{4}\,{\rm str} 
 \left( 
   u_\mu u^\mu + \chi_+  
 \right) \nonumber \\ 
 && 
 + e^2\,C\, 
   {\rm str}\left( 
             \widetilde{\mathcal{Q}}_L\widetilde{\mathcal{Q}}_R 
            \right)  
 - \frac{1}{4}\,F_{\mu\nu} F^{\mu\nu} 
 - \frac{\lambda}{2} \left(\del_\mu A^\mu\right)^2 \, . 
  \label{eq:chpt_2}
\end{eqnarray} 
 Using such a normalization of the decay constant 
$F_0$ that $F_0 \simeq 90\,{\rm MeV}$, 
$\Pi$ contains $\pi^+$, $\pi^3 /\sqrt{2}$ in 
its matrix element. 
 The coefficient $C$ parametrizes the EM correction 
induced from the short distance dynamics 
less than the length scale $\sim 1/\mu$  
above which the present effective description is valid.
 Since we adopt QED$_{L}$ constructed 
in Sec.~\ref{sec:finiteVolumeQED} in case space is finite,  
the gauge potential $A_\mu$ appearing 
in Eq.~(\ref{eq:chpt_2}) does not 
possess the components $\widetilde{A}_\mu(t,\,{\bf 0})$. 

 The (bare) intrinsic parity even Lagrangian density 
at the next-to-leading order  
consists of $\mathcal{O}(p^4)$-, 
$\mathcal{O}(e^2 p^2)$- and $\mathcal{O}(e^4)$-terms 
\begin{eqnarray} 
 \mathcal{L}_{4} 
 &=& 
 \sum_{j=0}^{12} L_j\,\mathfrak{X}_j 
 + 
 \sum_{j=1}^{25} e^2 F_0^2\,K_j\,\mathfrak{Y}_j \, . 
\end{eqnarray} 
 Here, $\mathfrak{X}_j$ ($0 = 1,\,\cdots,\,12$) 
are the terms of $\mathcal{O}(p^4)$  
\begin{eqnarray} 
 \mathfrak{X}_0 &\equiv& 
 {\rm str}\left(u_\mu u_\nu u^\mu u^\nu\right) \, , \nonumber \\ 
 \mathfrak{X}_1 &\equiv& 
 \left( 
  {\rm str}\left(u_\mu u^\mu\right) 
 \right)^2 \, , \nonumber \\ 
 \mathfrak{X}_2 &\equiv& 
 {\rm str}\left(u_\mu u_\nu\right) 
 {\rm str}\left(u^\mu u^\nu\right) \, , \nonumber \\ 
 \mathfrak{X}_3 &\equiv& 
 {\rm str} 
 \left( 
  \left(u_\mu u^\mu\right)^2 
 \right) \, , \nonumber \\ 
 \mathfrak{X}_4 &\equiv& 
 {\rm str}\left(u_\mu u^\mu\right)\, 
 {\rm str}\left(\chi_+\right) \, , \nonumber \\ 
 \mathfrak{X}_5 &\equiv& 
 {\rm str}\left(u_\mu u^\mu \chi_+\right) \, , \nonumber \\ 
 \mathfrak{X}_6 &\equiv& 
 \left({\rm str}\left(\chi_+\right)\right)^2 \, , \nonumber \\ 
 \mathfrak{X}_7 &\equiv& 
 \left({\rm str}\left(\chi_-\right)\right)^2 \, , \nonumber \\ 
 \mathfrak{X}_8 &\equiv& 
 \frac{1}{2}\,
 {\rm str}\left(\chi_+^2 + \chi_-^2\right) \, , \nonumber \\ 
 \mathfrak{X}_{9} &\equiv& 
 - \frac{i}{2}\, 
 {\rm str} 
 \left( 
  \left[u^\mu,\,u^\nu\right] 
  \mathfrak{F}_{+,\,\mu\nu} 
 \right) \, , \nonumber \\ 
 \mathfrak{X}_{10} &\equiv& 
 \frac{1}{4}\,
 {\rm str} 
 \left( 
  \mathfrak{F}_{+,\,\mu\nu} \mathfrak{F}_+^{\mu\nu} 
  - 
  \mathfrak{F}_{-,\,\mu\nu} \mathfrak{F}_-^{\mu\nu} 
 \right) \, , 
  \nonumber \\ 
 \mathfrak{X}_{11} 
 &\equiv& 
 {\rm str} 
 \left( 
  \mathcal{R}_{\mu\nu} \mathcal{R}^{\mu\nu} 
  + 
  \mathcal{L}_{\mu\nu} \mathcal{L}^{\mu\nu} 
 \right) \nonumber \\ 
 \mathfrak{X}_{12} &\equiv& 
 \frac{1}{4}\, 
 {\rm str}\left(\chi_+^2 - \chi_-^2\right) 
 = 
 {\rm str}\left(\chi \chi^\dagger\right) \, . 
  \label{eq:defOfX}
\end{eqnarray} 
 $L_{11,\,12}$ are known as $H_1 = L_{11}$ $H_2 = L_{12}$ 
\cite{Gasser:1984gg}. 
 Under the condition (\ref{eq:ConditionOnLRChargeMatrix}) 
and the leading-order equations of motion, 
$\mathfrak{Y}_j$ ($j=1,\,\cdots,\,25$) 
describe all possible EM corrections at $O(e^2 p^2)$ or $O(e^4)$ 
in terms of Nambu-Goldstone boson fields. 
 The first 19 ones are as follows;  
\begin{eqnarray} 
 \mathfrak{Y}_1 &=& 
 \frac{1}{2}\, 
 {\rm str} 
 \left( 
  \left(\widetilde{\mathcal{Q}}_L\right)^2 
  + 
  \left(\widetilde{\mathcal{Q}}_R\right)^2 
 \right) 
 {\rm str} 
 \left( 
  u_\mu u^\mu 
 \right) \, , \nonumber \\ 
 \mathfrak{Y}_2 &=& 
 {\rm str} 
 \left( 
  \widetilde{\mathcal{Q}}_L \widetilde{\mathcal{Q}}_R 
 \right) 
 {\rm str} 
 \left( 
  u_\mu u^\mu 
 \right) \, , \nonumber \\ 
 \mathfrak{Y}_3 &=& 
 - 
 {\rm str} 
 \left( 
  \widetilde{\mathcal{Q}}_R u_\mu   
 \right) 
 {\rm str} 
 \left( 
  \widetilde{\mathcal{Q}}_R u^\mu  
 \right) 
 - 
 {\rm str} 
 \left( 
  \widetilde{\mathcal{Q}}_L u_\mu   
 \right) 
 {\rm str} 
 \left( 
  \widetilde{\mathcal{Q}}_L u^\mu  
 \right) 
 \, , \nonumber \\ 
 \mathfrak{Y}_4 &=& 
 {\rm str} 
 \left( 
  \widetilde{\mathcal{Q}}_R u_\mu   
 \right) 
 {\rm str} 
 \left( 
  \widetilde{\mathcal{Q}}_L u^\mu  
 \right) \, , \nonumber \\ 
 \mathfrak{Y}_5 &=& 
 {\rm str} 
 \left[
  \left\{ 
   \left(\widetilde{\mathcal{Q}}_L\right)^2 
   + 
   \left(\widetilde{\mathcal{Q}}_R\right)^2 
  \right\} u_\mu u^\mu 
 \right] \, , \nonumber \\ 
 \mathfrak{Y}_6 &=& 
 {\rm str} 
 \left(
  \left( 
   \widetilde{\mathcal{Q}}_R \widetilde{\mathcal{Q}}_L 
   + 
   \widetilde{\mathcal{Q}}_L \widetilde{\mathcal{Q}}_R 
  \right) u_\mu u^\mu 
 \right) \, , \nonumber \\ 
 \mathfrak{Y}_7 &=& 
 \frac{1}{2}\, 
 {\rm str} 
 \left( 
  \left(\widetilde{\mathcal{Q}}_L\right)^2 
  + 
  \left(\widetilde{\mathcal{Q}}_R\right)^2 
 \right) 
 {\rm str}\left(\chi_+\right) \, , \nonumber \\ 
 \mathfrak{Y}_8 &=& 
 {\rm str} 
 \left( 
  \widetilde{\mathcal{Q}}_L \widetilde{\mathcal{Q}}_R 
 \right) 
 {\rm str}\left(\chi_+\right) \, , \nonumber \\ 
 \mathfrak{Y}_9 &=& 
 {\rm str} 
 \left[
  \left\{ 
   \left(\widetilde{\mathcal{Q}}_L\right)^2 
   + 
   \left(\widetilde{\mathcal{Q}}_R\right)^2 
  \right\} \chi_+ 
 \right] \, , \nonumber \\ 
 \mathfrak{Y}_{10} &=& 
 {\rm str} 
 \left(
  \left( 
   \widetilde{\mathcal{Q}}_R \widetilde{\mathcal{Q}}_L 
   + 
   \widetilde{\mathcal{Q}}_L \widetilde{\mathcal{Q}}_R 
  \right) \chi_+ 
 \right) \, , \nonumber \\ 
 \mathfrak{Y}_{11} &=& 
 {\rm str} 
 \left(
  \left( 
   \widetilde{\mathcal{Q}}_R \widetilde{\mathcal{Q}}_L 
   - 
   \widetilde{\mathcal{Q}}_L \widetilde{\mathcal{Q}}_R 
  \right) \chi_- 
 \right) \, , \nonumber \\ 
 \mathfrak{Y}_{12} &=& 
 i\, 
 {\rm str} 
 \left(
  \left[ 
   \nabla_\mu \widetilde{\mathcal{Q}}_R ,\, 
   \widetilde{\mathcal{Q}}_R 
  \right] u^\mu 
  - 
  \left[ 
   \nabla_\mu \widetilde{\mathcal{Q}}_L ,\, 
   \widetilde{\mathcal{Q}}_L 
  \right] u^\mu 
 \right) \, , \nonumber \\ 
 \mathfrak{Y}_{13} &=& 
 {\rm str} 
 \left( 
  \nabla_\mu \widetilde{\mathcal{Q}}_R 
  \nabla^\mu \widetilde{\mathcal{Q}}_L 
 \right) \, , \nonumber \\ 
 \mathfrak{Y}_{14} &=& 
 {\rm str} 
 \left( 
  \nabla_\mu \widetilde{\mathcal{Q}}_R 
  \nabla^\mu \widetilde{\mathcal{Q}}_R 
  + 
  \nabla_\mu \widetilde{\mathcal{Q}}_L 
  \nabla^\mu \widetilde{\mathcal{Q}}_L 
 \right) \, , \nonumber \\ 
 \mathfrak{Y}_{15} &=& 
 e^2 F_0^2 
 \left( 
  {\rm str} 
  \left( 
   \widetilde{\mathcal{Q}}_R \widetilde{\mathcal{Q}}_L 
  \right) 
 \right)^2 \, , \nonumber \\ 
 \mathfrak{Y}_{16} &=& 
 e^2 F_0^2\, 
 {\rm str} 
 \left( 
  \widetilde{\mathcal{Q}}_R \widetilde{\mathcal{Q}}_L 
 \right) 
 {\rm str} 
 \left( 
  \left(\widetilde{\mathcal{Q}}_R\right)^2 
  + 
  \left(\widetilde{\mathcal{Q}}_L\right)^2 
 \right) \, , \nonumber \\ 
 \mathfrak{Y}_{17} &=& 
 e^2 F_0^2 
 \left( 
  {\rm str} 
  \left( 
   \left(\widetilde{\mathcal{Q}}_R\right)^2 
   + 
   \left(\widetilde{\mathcal{Q}}_L\right)^2 
  \right) 
 \right)^2 \, , \nonumber \\ 
 \mathfrak{Y}_{18} &=& 
 {\rm str} 
 \left( 
  \widetilde{\mathcal{Q}}_R u_\mu 
  \widetilde{\mathcal{Q}}_R u^\mu 
  + 
  \widetilde{\mathcal{Q}}_L u_\mu 
  \widetilde{\mathcal{Q}}_L u^\mu 
 \right) \, , \nonumber \\ 
 \mathfrak{Y}_{19} &=& 
 {\rm str} 
 \left( 
  \widetilde{\mathcal{Q}}_R u_\mu 
  \widetilde{\mathcal{Q}}_L u^\mu 
 \right) \, . \label{eq:Y_1}
\end{eqnarray} 
 The rest 6 ones are as follows; 
\begin{eqnarray} 
 \mathfrak{Y}_{20} &=& 
 e^2 F_0^2\, 
 {\rm str} 
 \left( 
  \left(\widetilde{\mathcal{Q}}_R\right)^2 
  \left(\widetilde{\mathcal{Q}}_L\right)^2  
 \right) \, , \nonumber \\ 
 \mathfrak{Y}_{21} &=& 
 e^2 F_0^2\, 
 {\rm str} 
 \left( 
  \widetilde{\mathcal{Q}}_R \widetilde{\mathcal{Q}}_L  
  \widetilde{\mathcal{Q}}_R \widetilde{\mathcal{Q}}_L  
 \right) \, , \nonumber \\ 
 \mathfrak{Y}_{22} &=& 
 e^2 F_0^2 
 \left\{  
  {\rm str} 
  \left( 
   \left(\widetilde{\mathcal{Q}}_R\right)^2 
   - 
   \left(\widetilde{\mathcal{Q}}_L\right)^2 
  \right) 
 \right\}^2 \, , \nonumber \\ 
 \mathfrak{Y}_{23} &=& 
 e^2 F_0^2 
 \left\{ 
  {\rm str}\left(\widetilde{\mathcal{Q}}_R\right) 
  {\rm str}
  \left( 
   \widetilde{\mathcal{Q}}_R
   \left(\widetilde{\mathcal{Q}}_L\right)^2 
  \right) 
  + 
  {\rm str}\left(\widetilde{\mathcal{Q}}_L\right) 
  {\rm str}
  \left( 
   \widetilde{\mathcal{Q}}_L
   \left(\widetilde{\mathcal{Q}}_R\right)^2 
  \right) 
 \right\} \, , \nonumber \\ 
 \mathfrak{Y}_{24} &=& 
 {\rm str}\left(\widetilde{\mathcal{Q}}_R\right) 
 {\rm str}
 \left( 
  \widetilde{\mathcal{Q}}_L u_\mu u^\mu 
 \right) 
 + 
 {\rm str}\left(\widetilde{\mathcal{Q}}_L\right) 
 {\rm str}
 \left( 
  \widetilde{\mathcal{Q}}_R u_\mu u^\mu 
 \right) \, , \nonumber \\ 
 \mathfrak{Y}_{25} &=& 
 {\rm str}\left(\widetilde{\mathcal{Q}}_R\right) 
 {\rm str}
 \left( 
  \widetilde{\mathcal{Q}}_L \chi_+  
 \right) 
 + 
 {\rm str}\left(\widetilde{\mathcal{Q}}_L\right) 
 {\rm str}
 \left( 
  \widetilde{\mathcal{Q}}_R \chi_+ 
 \right) \, . \label{eq:Y_2}
\end{eqnarray} 
\TABLE[p]{ 
\begin{tabular}{|c|c||c|c|} 
\hline
 $j$ & $k_j$ & $j$ & $k_j$ \\ 
\hline
 $1$ & $0$ & $14$ & $0$ \\ 
 $2$ & $\mathcal{Z}$ & $15$ & $\frac{3}{2} + 8\,\mathcal{Z}^2$ \\ 
 $3$ & $0$ & $16$ & $-\frac{3}{2}$ \\ 
 $4$ & $2\,\mathcal{Z}$ & $17$ & $\frac{3}{8}$ \\ 
 $5$ & $-\frac{3}{4}$ & $18$ & $\frac{3}{4}$ \\ 
 $6$ & $\frac{N_S}{2}\,\mathcal{Z}$ & $19$ & $0$ \\ 
 $7$ & $0$ & $20$ & $2 N_S \mathcal{Z}^2 - 3 \mathcal{Z}$ \\ 
 $8$ & $\mathcal{Z}$ & 
   $21$ & $2 N_S \mathcal{Z}^2 + 3 \mathcal{Z}$ \\  
 $9$ & $-\frac{1}{4}$ & $22$ & $-\mathcal{Z}^2$ \\  
 $10$ & $\frac{1}{4} + \frac{N_S}{2}\,\mathcal{Z}$ & 
  $23$ & $-8 \mathcal{Z}^2$ \\ 
 $11$ & $-\frac{1}{8}$ & $24$ & $- \mathcal{Z}$ \\ 
 $12$ & $\frac{1}{4}$ & $25$ & $- \mathcal{Z}$ \\ 
 $13$ & $0$ & & \\ 
 \hline 
\end{tabular} 
\caption{Coefficients $k_j$ of UV divergence in $K_j$.} 
\label{tab:k_j}
} 

 The coefficients $L_j$, $K_k$ absorb the UV divergences 
that arise from the one-loop correction 
\begin{eqnarray} 
 & 
 \displaystyle{ 
  L_j + l_j\,\Delta_\epsilon = L^{\bf R}_j(\mu) \, , 
 }& \nonumber \\ 
 & 
 \displaystyle{ 
  K_j + k_j\,\Delta_\epsilon = K^{\bf R}_j(\mu) \, . 
 }& \label{eq:renormalizationOfLandK}
\end{eqnarray} 
 Here we employ the dimensional regularization where 
the spatial dimension is analytically continued to $d$. 
 Denoting the full space-time dimension 
as $D = d + 1 = 4 - 2 \epsilon$, 
$\Delta_\epsilon$ in Eq.~(\ref{eq:renormalizationOfLandK}) is 
given by 
\begin{eqnarray} 
 \Delta_\epsilon &\equiv& 
   \frac{1}{32 \pi^2} 
   \left\{ 
    \frac{1}{\epsilon} 
    - {\rm ln}\left(\frac{\mu^2}{4\pi}\right) - \gamma_E + 1 
   \right\} \, . \nonumber \\ 
 \label{eq:renormalizationLEC}
\end{eqnarray} 
 The values of $l_j$ ($j = 1,\,\cdots,\,12$) 
are available in Ref.~\cite{Bijnens:1999hw} 
for the generic number of flavors.  
 The values of $k_j$ ($j = 1,\,\cdots,\,14$, 
\,$18$,\,$19$)  
were computed in Ref.~\cite{Bijnens:2006mk} 
for $N_S = 3$ . 
 Table \ref{tab:k_j} lists the values of $k_j$ 
\footnote{ 
 The normalization of our $k_j$'s differs 
from those in Ref.~\cite{Bijnens:2006mk} such that  
$(k_j)_{\rm ours} = -2 (k_j)_{\rm BD}$ 
for $j=1,\,\cdots,\,10,\,12,\,13,\,14,\,18,\,19$, 
and $(k_{11})_{\rm ours} = 2 (k_{11})_{\rm BD}$. 
} 
computed using the heat kernel method 
and evaluating the fermionic pion contribution 
explicitly for general $N_S$ without setting the quantities 
in Eq.~(\ref{eq:ConditionOnLRChargeMatrix}) to zero. 
 The compuatation is performed only for Feynman gauge $\lambda = 1$. 
 The dimensionless quantity $\mathcal{Z}$ 
in Table \ref{tab:k_j} is defined by 
\begin{eqnarray} 
 \mathcal{Z} \equiv \frac{C}{F_0^4} \, . 
\end{eqnarray} 

 For the sake of simplifying expressions, 
the superscript ``{\bf R}'' is omitted 
from every renormalized parameter 
that appears in what follows. 

 In the momentum space, 
the Feynman rule in the partially quenched chiral perturbation theory 
in finite volume is the same as that in infinite volume. 
 In particular, the form of propagators 
that allows us to carry out the computation efficiently 
is obtained by introducing the super-traceless component 
called super-$\eta^\prime$, 
deriving the propagators in the matrix element basis 
$\Phi^I_{\ J}$ ($I,\,J = 1,\,\cdots,\,(N_S + 2 N_V)$),  
and taking the decoupling limit of super-$\eta^\prime$ 
\cite{Sharpe:2000bc,Sharpe:2001fh}. 

 The entry $\chi_{IJ}$ of the matrix of meson mass squared  
including the leading-order EM correction is given by 
\begin{eqnarray} 
 \chi_{IJ} &=& 
 \frac{\chi_I + \chi_J}{2} 
 + \frac{2 e^2 C}{F_0^2} \left(q_I - q_J\right)^2 \, , 
  \label{eq:DefOfChi}
\end{eqnarray} 
where $\chi_I$ and $q_I$ are the eigenvalues 
of $\left.\chi\right|_{\mathcal{M} \rightarrow M_d}$ 
and $\mathcal{Q}$ in Eq.~(\ref{eq:DiagonalchargeMatrix}), 
respectively. 
 The propagators are shown here  
in the case that the eigenvalues $\chi_{(r)} \equiv \chi_{N_V + r}$ 
($1 \le r \le N_S$) corresponding to sea quarks 
as well as the mass squared 
$\chi_x$ ($x = 1,\,\cdots,\,N_S - 1$) 
of the diagonal meson eigenstates 
in the sea-meson sub-sector,   
referred to as {\it sea mesons}, 
differ from all $\chi_j$ ($j = 1,\,\cdots,\,N_V$). 
 The propagators of bosonic mesons 
($1 \le I,\,J,\,K,\,L \le (N_V + N_S)$ 
or $(1 + N_V + N_S) \le I,\,J,\,K,\,L \le (2 N_V + N_S)$) 
are denoted by 
\begin{eqnarray} 
 i\,G^{I\ \ \,K}_{\ J\,;\ \ L}(p^2) 
 &\equiv&  
 \int d^4 x\,e^{i p \cdot x}\, 
 \left<\Pi^I_{\ J}(x)\,\Pi^K_{\ L}(0)\right> \, . 
\end{eqnarray} 
 The off-diagonal meson fields have the usual 
form of propagators 
\begin{eqnarray} 
 i\,G^{I\ \ \,K}_{\ J\,;\ \ L}(p^2) 
 &=& 
 \delta^I_{\ L}\,\delta_J^{\ K}\,\frac{i}{p^2 - \chi_{IJ}} 
 \quad (I \ne J,\,K \ne L) \, . 
\end{eqnarray} 
 The propagators of the fermionic mesons 
$\Xi^i_{\ j} \equiv \Pi^{i + N_S + N_V}_{\qquad \quad \ j}$ 
($1 \le i \le N_V$,\,$1 \le j \le (N_S + N_V)$) 
also have simple forms 
\begin{eqnarray} 
 i S^{i\ \ \,k}_{\ j;\,\ \ l}(p^2) 
 &\equiv& 
 \int d^4 x\,e^{i p \cdot x}\, 
 \left<\Xi^i_{\ j}(x)\,\Xi^{\dagger\,k}_{\ \ l}(0)\right> 
  \nonumber \\ 
 &=& 
 \delta^i_{\ l}\,\delta_j^{\ k}\,\frac{i}{p^2 - \chi_{ij}} 
 \quad 
 (1 \le i,\,l \le N_V,\,1 \le j,\,k \le (N_S + N_V)) \, .  
\end{eqnarray} 
 The one-loop calculation needs the propagators of diagonal mesons 
only for $i,\,j = 1,\,\cdots,\,N_V$;  
\begin{eqnarray} 
 G^{i\ \ \,j}_{\ i\,;\ \ j}(p^2) 
 &=& 
 - \frac{1}{N_S} 
 \left( 
  \frac{R_{ij}^{\ \ i}}{p^2 - \chi_i} 
  + 
  \frac{R_{ij}^{\ \ j}}{p^2 - \chi_j} 
  + 
  {\sum_{x}}^{{\rm sm}} 
   \frac{R_{ij}^{\ \ x}}{p^2 - \chi_x} 
 \right) 
 \quad {\rm for}\ \chi_i \ne \chi_j \, , \nonumber \\ 
 G^{i\ \ \,j}_{\ i\,;\ \ j}(p^2) 
 &=& 
 \delta_{ij}\,\frac{1}{p^2 - \chi_i} 
  \nonumber \\ 
 && 
 - \frac{1}{N_S} 
 \left( 
  \frac{R_i^{(d)}}{(p^2 - \chi_i)^2} 
  + 
  \frac{R_i^{(s)}}{p^2 - \chi_i} 
  + 
  {\sum_{x}}^{{\rm sm}} 
   \frac{R_{ii}^{\ \ x}}{p^2 - \chi_x} 
 \right) \nonumber \\ 
 && \qquad 
 {\rm for}\ \chi_i = \chi_j \, ,  
\end{eqnarray} 
where $\displaystyle{{\sum_x}^{\rm sm}}$ 
denotes the sum over all sea mesons, and 
\begin{eqnarray} 
 R_{ij}^{\ \ i} 
 &\equiv& 
 \frac{ 
    \displaystyle{ 
     \prod_{r=1}^{N_S} \left(\chi_i - \chi_{(r)}\right) 
    }  
  } 
  { 
    \displaystyle{ 
     \left(\chi_i - \chi_j\right) 
     {\prod_x}^{\rm sm} \left(\chi_i - \chi_x\right)
    }  
  } 
  = R_{ji}^{\ \ i} \, , \nonumber \\ 
%
 R_{ij}^{\ \ x} 
 &\equiv& 
 \frac{ 
  \displaystyle{ 
   \prod_{r=1}^{N_S} \left(\chi_x - \chi_{(r)}\right) 
  }}
  {\displaystyle{  
    \left(\chi_x - \chi_i\right) \left(\chi_x - \chi_j\right) 
     {\prod_{y \ne x}}^{\rm sm} \left(\chi_x - \chi_y\right)} 
  } \, , \nonumber \\ 
 R_i^{(d)} &\equiv& 
 \frac{ 
  \displaystyle{ 
   \prod_{r=1}^{N_S} \left(\chi_i - \chi_{(r)}\right) 
  }} 
  {\displaystyle{ 
    {\prod_x}^{\rm sm} \left(\chi_i - \chi_x\right) 
   }} \, , \nonumber \\ 
 R_i^{(s)} &=& 
 \frac{ 
  \displaystyle{ 
   \prod_{r=1}^{N_S} \left(\chi_i - \chi_{(r)}\right) 
  }} 
  {\displaystyle{ 
    {\prod_x}^{\rm sm} \left(\chi_i - \chi_x\right) 
   }}  
  \left( 
   \sum_{s=1}^{N_S} \frac{1}{\chi_i - \chi_{(s)}} 
   - {\sum_x}^{\rm sm} \frac{1}{\chi_i - \chi_x}     
  \right) \, . 
\end{eqnarray} 

\subsection{electromagnetic correction to meson mass 
in finite volume} 
\label{subsec:EMfiniteSizeCorrection} 

 With the preparations done in the previous subsection, 
we are ready to compute 
the next-to-leading order correction 
$\left.m_{ij}^2\right|_4(L)$ to 
the off-diagonal pseudoscalar meson mass squared 
in finite volume, which can be obtained from 
the self-energy function 
$\left.\Sigma^i_{\ j}(p^2)\right|_L$ 
within our approximation as 
\begin{eqnarray} 
 &  
 \displaystyle{ 
  \left.m_{ij}^2\right|_4(L) 
   = \left.\Sigma^i_{\ j}(\chi_{ij})\right|_L \, , 
 }& 
\end{eqnarray} 
and investigate the relevance of finite size correction 
to the EM splitting. 
 As recalled in Sec.~\ref{subsec:newQED} 
Lorentz boost symmetry is violated in finite volume. 
 Here $\left.m_{ij}^2\right|_4(L)$ 
is defined in the rest frame, 
$p^\mu = (\sqrt{\chi_{ij}},\,{\bf 0})$. 

 Before carrying on the explicit calculation further, 
we mention the limitation inherent to our approach 
on the ability to capture the dynamics. 
 While the low energy effective theory enables us to evaluate 
the effect of finiteness of volume 
on the virtual quanta with low frequencies, 
it provides no knowledge on the effect to the short distance dynamics 
packed in the low energy constants $L_j(\mu)$ and $K_j(\mu)$. 
 The finite size scaling effect 
on the low energy constants $L_j(\mu)$ 
carrying the information on QCD less than $1/\mu$,  
is insignificant so as to affect to the properties 
of Nambu-Goldstone bosons. 
 The situation is different in QED 
where there is no intrinsic scale $1 /\mu$ 
that separates long and short distances \cite{Gasser:2003hk}. 
 For this reason, though the finite size scaling effect on QED 
is dominated by the loop contribution whose finite part 
represents the long distance physics, the low energy constants 
$K_j(\mu)$ possibly suffer from sub-dominant 
but non-negligible finite size scaling effect. 
 For the purpose explained in Sec.~\ref{sec:introduction}, 
even qualitative features of the finite size scaling effect 
such as the size and sign are worthwhile to investigate. 
 
 With the above remark in mind, we proceed to 
calculate $\left.m_{ij}^2\right|_4(L)$ in finite volume. 
 For the purpose of 
(1) getting $\left.m^2_{ij}\right|_4(L)$ 
for $N_S = 2,\,3$ simultaneously, 
and (2) demonstrating 
that QED$_{L}$ share the common UV divergent structure 
as QED in infinite volume, 
we describe the computation in some detail. 
 As the probes are valence quarks, 
it suffices to put $1 \le i (\ne) j \le N_F$.  
 Straightforward calculation of four types of Feynman diagrams yields 
\begin{eqnarray} 
 \left.m_{ij}^2\right|_4(L) 
 &=& 
 \frac{1}{6 F_0^2} 
 \int_{-\infty}^\infty \frac{dk^0}{2\pi} 
 \frac{1}{L^d} \sum_{{\bf k} \in \widetilde{\Gamma}_d} 
 \left[ 
  2\,\left(k^2 + 2\,\chi^{\rm QCD}_{ij}\right) 
   G^{i\ \ j}_{\ i;\ \ j}(k^2) 
 \right. \nonumber \\ 
 && \qquad \qquad \quad 
  + 
  \left(\chi_i - k^2\right)    
  \left( 
   G^{i\ \ i}_{\ i;\ \ i}(k^2) - S^{i\ \ i}_{\ i;\ \ i}(k^2) 
  \right) \nonumber \\ 
 && \qquad \qquad \quad 
 \left. 
  + 
  \left(\chi_j - k^2\right)    
  \left( 
   G^{j\ \ j}_{\ j;\ \ j}(k^2) - S^{j\ \ j}_{\ j;\ \ j}(k^2) 
  \right) 
 \right] \nonumber \\ 
 && \   
  + 2\,e^2 \mathcal{Z}  
  {\sum_n}^S 
  \int_{-\infty}^\infty \frac{dk^0}{2\pi} 
  \frac{1}{L^d} \sum_{{\bf k} \in \widetilde{\Gamma}_d} 
  \left\{ 
   q_{ij} q_{in}\,G^{i\ \ n}_{\ n;\ \ i}(k^2) 
  \right. \nonumber \\ 
 && \qquad \qquad \qquad \qquad \qquad \qquad \quad  
  \left. 
   + 
   q_{ij} q_{nj}\,G^{n\ \ j}_{\ j;\ \ n}(k^2) 
  \right\} \nonumber \\ 
 && \ 
 + 
 \left(q_{ij}\right)^2 e^2 
 \int_{-\infty}^\infty \frac{dk^0}{2\pi} 
 \frac{1}{L^d} \sum_{{\bf k} \in \widetilde{\Gamma}_d^\prime} 
 \left\{ 
  \left(D - 1\right) 
  \frac{1}{-k^2} 
 \right. \nonumber \\ 
 && \qquad \qquad \qquad \qquad 
  + 
  2\, 
  \frac{p \cdot k}{-k^2 \left(\chi_{ij} - (k+p)^2\right)}  
   \nonumber \\ 
 && \qquad \qquad \qquad \qquad 
 \left. 
  + 
  4\,\chi_{ij}\, 
  \frac{1}{-k^2 \left(\chi_{ij} - (k+p)^2\right)}  
 \right\} \nonumber \\ 
 && 
 + m_{C,\,ij}^2\, , 
  \label{eq:totalOneLoop}
\end{eqnarray} 
where $\displaystyle{{\sum_n}^S}$ represents the sum over 
the indices $n$ of sea quark flavors only, 
$\widetilde{\Gamma}_d^\prime 
\equiv \widetilde{\Gamma}_d - \left\{{\bf 0}\right\}$ 
with $\widetilde{\Gamma}_d$ in Eq.~(\ref{eq:def_tildeGamma_d}), 
and 
\begin{eqnarray} 
 \chi^{\rm QCD}_{ij} &\equiv& \frac{\chi_i + \chi_j}{2} \, , 
  \nonumber \\ 
 q_{ij} &\equiv& q_i - q_j \, .   
\end{eqnarray} 
 $m_{C,\,ij}^2$ represents the contribution 
of the next-to-leading order local terms in 
Eqs.~(\ref{eq:defOfX}), (\ref{eq:Y_1}) and (\ref{eq:Y_2})
to the pseudoscalar meson mass squared. 
 $m_{C,\,ij}^2$ in Eq.~(\ref{eq:totalOneLoop}) 
is written in terms of bare $L_j$, $K_j$. 
 The explicit form of $m_{C,\,ij}^2$ 
will be given after renormalization is performed 
(the sum of $\left.m^2_{{\rm C},\,ij}\right|^{\rm QCD}$ 
and $\left.m^2_{{\rm C},\,ij}\right|^{\rm EM}$ 
in Eq.~(\ref{eq:LECcontribution})).  
 We note that the expression 
in Eq.~(\ref{eq:totalOneLoop}) is independent 
of the values of the gauge parameter $\lambda$. 

 Eq.~(\ref{eq:totalOneLoop}) can be rewritten in terms 
of four basic functions,  
(\ref{eq:sum_I11}), (\ref{eq:sum_J11}), (\ref{eq:sum_I_1-0}) 
and (\ref{eq:sum_I_1}), 
defined in Appendix \ref{sec:basicSums}. 
 The function $I_1(m^2;\,L)$ in Eq.~(\ref{eq:sum_I_1}) 
emerges in the QCD finite scaling effect 
\cite{AliKhan:2003cu}. 
 In Appendix \ref{sec:basicSums}, 
we derive the expression written in terms of 
a Jacobi theta function for the other three functions. 
 Applying the identities 
\begin{eqnarray} 
 & 
 \displaystyle{ 
  \left(2\chi_i + \chi_j\right) R_{ij}^{\ \ i} 
  - \frac{1}{2}\,R^{(d)}_i 
  = \frac{3}{2} \left(\chi_i + \chi_j\right) R_{ij}^{\ \ i} \, , 
 }& \nonumber \\ 
 & 
 \displaystyle{ 
  \left(\chi_i + 2\,\chi_j\right) R_{ij}^{\ \ j} 
  - \frac{1}{2}\,R^{(d)}_j 
  = \frac{3}{2} \left(\chi_i + \chi_j\right) R_{ij}^{\ \ j} \, , 
 }& \nonumber \\ 
 & 
 \displaystyle{ 
  \left(\chi_x + \chi_i + \chi_j\right) R_{ij}^{\ \ x} 
   + \frac{\chi_i - \chi_x}{2}\,R_{ii}^{\ \ x} 
   + \frac{\chi_j - \chi_x}{2}\,R_{jj}^{\ \ x} 
  = \frac{3}{2} \left(\chi_i + \chi_j\right) R_{ij}^{\ \ x} \, , 
 }& 
\end{eqnarray} 
to the terms in Eq.~(\ref{eq:totalOneLoop}) 
that are written solely by $I_1(m^2;\,L)$, 
we get a compact expression  
\begin{eqnarray} 
 \left.m_{ij}^2\right|_4(L)   
 &=& 
 \frac{1}{2 N_S F_0^2} \left(\chi_i + \chi_j\right) 
  \nonumber \\ 
 && \  
 \times 
 \left\{ 
  R_{ij}^{\ \ i}\,I_1(\chi_i;\,L) 
  + R_{ij}^{\ \ j}\,I_1(\chi_j;\,L) 
  + 
  {\sum_x}^{\rm sm} 
   R_{ij}^{\ \ x} I_1(\chi_x;\,L) 
 \right\} \nonumber \\ 
 && 
 - 
 2 e^2 \mathcal{Z} {\sum_n}^{\rm S} 
 \left\{ 
  q_{ij} q_{in} I_1(\chi_{in};\,L) 
  + q_{ij} q_{nj} I_1(\chi_{nj};\,L)
 \right\} \nonumber \\ 
 && 
 + 
 \left(q_{ij}\right)^2 e^2 
 \left\{ 
  \left(D-1\right) I_1^0(L) 
  + 2\,J_{11}(\chi_{ij};\,L) 
  + 4\,\chi_{ij}\,I_{11}(\chi_{ij};\,L) 
 \right\} \nonumber \\ 
 && 
 + m_{C,\,ij}^2 \, . 
  \label{eq:keyExpression}
\end{eqnarray} 

 From Eq.~(\ref{eq:keyExpression}), 
the finite size scaling correction is obtained as 
\begin{eqnarray} 
 \Delta m_{ij}^2(L) 
 &=& 
 \left.m_{ij}^2\right|_4(L) - \left.m_{ij}^2\right|_4(\infty) 
  \, , 
\end{eqnarray} 
using the quantity $\left.m_{ij}^2\right|_4(\infty)$ 
evaluated directly in infinite volume. 
 The use of the results in Appendix \ref{sec:basicSums} 
immediately leads  
\begin{eqnarray} 
 \Delta m_{ij}^2(L) 
 &=& 
 \Delta m^2_{{\rm QCD},\,ij}(L) + \Delta m^2_{{\rm EM},\,ij}(L) \, , 
\end{eqnarray} 
where $\Delta m^2_{{\rm QCD},\,ij}(L)$ 
and $\Delta m^2_{{\rm EM},\,ij}(L)$ 
represent the finite size scaling correction 
on QCD and QED, respectively 
\begin{eqnarray} 
 \Delta m^2_{{\rm QCD},\,ij}(L) 
 &=& 
 \frac{1}{\left(4\pi\right)^2} 
 \frac{\chi_i + \chi_j}{2 N_S F_0^2}\, 
 \frac{1}{L^2}  
 \left\{ 
  R_{ij}^{\ \ i}\,\mathcal{M}(\sqrt{\chi_i} L) 
  + 
  R_{ij}^{\ \ j}\,\mathcal{M}(\sqrt{\chi_j} L) 
 \right. \nonumber \\ 
 && \qquad \qquad \qquad \qquad \quad 
 \left. 
  + 
  {\sum_x}^{\rm sm} 
   R_{ij}^{\ \ x}\,\mathcal{M}(\sqrt{\chi_x} L)     
 \right\} \, , \nonumber \\ 
 \Delta m^2_{{\rm EM},\,ij}(L) 
 &=& 
 - 
 \frac{2 e^2 \mathcal{Z}}{\left(4\pi\right)^2} 
 \frac{1}{L^2}  
 {\sum_{n}}^S 
 \left\{ 
  q_{ij}\,q_{in}\,\mathcal{M}(\sqrt{\chi_{in}} L) 
  + 
  q_{ij}\,q_{nj}\,\mathcal{M}(\sqrt{\chi_{nj}} L) 
 \right. \nonumber \\ 
 && 
 - 3\,\frac{\left(q_{ij}\right)^2 e^2}{4\pi}\,\frac{\kappa}{L^2} 
  \nonumber \\ 
 && 
 + 
 \frac{\left(q_{ij}\right)^2 e^2}{\left(4\pi\right)^2} 
 \left\{ 
    \frac{\mathcal{K}(\sqrt{\chi_{ij}} L)}{L^2} 
  - 4\,\sqrt{\chi_{ij}}\,\frac{\mathcal{H}(\sqrt{\chi_{ij}} L)}{L} 
 \right\} \, . \label{eq:finiteSizeCorrection}
\end{eqnarray} 
 This expression is free from UV divergence, 
confirming the statement at the end of Sec.~\ref{subsec:newQED}. 
 Carrying out the renormalization for 
the part in infinite volume with help 
of the values of $k_j$'s in Table \ref{tab:k_j}, 
we get the finite expression of 
the next-to-leading correction $\left.m^2_{ij}\right|_4(L)$ 
\begin{eqnarray} 
 \left.m^2_{ij}\right|_4(L) 
 &=& 
 \left.m^2_{{\rm QCD},\,ij}\right|_4(L) 
 + 
 \left.m^2_{{\rm EM},\,ij}\right|_4(L) \, , 
\end{eqnarray} 
with $\left.m^2_{{\rm QCD},\,ij}\right|_4(L)$ 
($\left.m^2_{{\rm EM},\,ij}\right|_4(L)$) 
composed of the corresponding quantity     
$\left.m^2_{{\rm QCD},\,ij}\right|_4(\infty)$ 
($\left.m^2_{{\rm EM},\,ij}\right|_4(\infty)$) 
in infinite volume and the finite size correction in 
Eq.~(\ref{eq:finiteSizeCorrection}) 
\begin{eqnarray} 
 \left.m^2_{{\rm QCD},\,ij}\right|_4(L) &=& 
 \left.m^2_{{\rm QCD},\,ij}\right|_4(\infty) + 
  \Delta m^2_{{\rm QCD},\,ij}(L) \, , 
  \nonumber \\ 
 \left.m^2_{{\rm EM},\,ij}\right|_4(L) &=& 
 \left.m^2_{{\rm EM},\,ij}\right|_4(\infty) 
  + \Delta m^2_{{\rm EM},\,ij}(L) \, . 
\end{eqnarray} 
 Each of $\left.m^2_{{\rm QCD},\,ij}\right|_4(\infty)$ 
and $\left.m^2_{{\rm EM},\,ij}\right|_4(\infty)$ consists 
of two parts 
\begin{eqnarray} 
 \left.m^2_{{\rm QCD},\,ij}\right|_4(\infty) 
 &=& 
 \left.m^2_{{\rm QCD},\,ij}(\infty)\right|^{\rm loop} 
 + 
 \left.m^2_{{\rm C},\,ij}\right|^{\rm QCD} \, , 
  \nonumber \\ 
 \left.m^2_{{\rm EM},\,ij}\right|_4(\infty) 
 &=& 
 \left.m^2_{{\rm EM},\,ij}(\infty)\right|^{\rm loop} 
 + 
 \left.m^2_{{\rm C},\,ij}\right|^{\rm EM} \, . 
\end{eqnarray} 
 The terms $\left.m^2_{{\rm QCD},\,ij}(\infty)\right|^{\rm loop}$, 
$\left.m^2_{{\rm EM},\,ij}(\infty)\right|^{\rm loop}$ 
are those involving the chiral logarithms 
\begin{eqnarray} 
 \left.m^2_{{\rm QCD},\,ij}(\infty)\right|^{\rm loop} 
 &=& 
 \frac{1}{\left(4\pi\right)^2} 
 \frac{\chi_i + \chi_j}{2 N_S F_0^2} 
 \left\{ 
  R_{ij}^{\ \ i}\,\chi_i\,{\rm ln}\left(\frac{\chi_i}{\mu^2}\right) 
  + 
  R_{ij}^{\ \ j}\,\chi_j\,{\rm ln}\left(\frac{\chi_j}{\mu^2}\right) 
 \right. \nonumber \\ 
 && \qquad \qquad \qquad \qquad \qquad 
 \left. 
  + 
  \sum_x 
  R_{ij}^{\ \ x}\,\chi_x\,{\rm ln}\left(\frac{\chi_x}{\mu^2}\right)
 \right\} \, , \nonumber \\ 
 \left.m^2_{{\rm EM},\,ij}(\infty)\right|^{\rm loop} 
 &=& 
 - 
 \frac{2 e^2 \mathcal{Z}}{\left(4\pi\right)^2} 
 {\sum_n}^{\rm S} 
 \left\{ 
  q_{ij} q_{in}\, 
  \chi_{in}\,{\rm ln}\left(\frac{\chi_{in}}{\mu^2}\right) 
  + 
  q_{ij} q_{nj}\,
  \chi_{nj}\,{\rm ln}\left(\frac{\chi_{nj}}{\mu^2}\right) 
 \right\} \nonumber \\ 
 && 
 -  
 \frac{\left(q_{ij}\right)^2 e^2}{\left(4\pi\right)^2}\, 
 \chi_{ij} 
 \left\{ 
  3\,{\rm ln}\left(\frac{\chi_{ij}}{\mu^2}\right) - 4 
 \right\} \, ,  
\end{eqnarray} 
while the terms 
$\left.m^2_{{\rm C},\,ij}\right|^{\rm QCD}$ 
and $\left.m^2_{{\rm C},\,ij}\right|^{\rm EM}$ 
are those given in terms of low energy constants 
\begin{eqnarray} 
 \left.m^2_{{\rm C},\,ij}\right|^{\rm QCD} 
 &=& 
 \frac{1}{F_0^2} 
 \left[ 
  - 4 \left\{ 
       2\,L_4\,N_S\,\overline{\chi}_S 
       + L_5 \left(\chi_i + \chi_j\right)
      \right\} \chi_{ij} 
 \right. \nonumber \\ 
 && \qquad 
 \left. 
  + 8\,L_6\,N_S\,\overline{\chi}_S \left(\chi_i + \chi_j\right) 
  + 4\,L_8 \left(\chi_i + \chi_j\right)^2 
 \right] \, , \nonumber \\ 
 \left.m^2_{{\rm C},\,ij}\right|^{\rm EM} 
 &=& 
 e^2 
 \left[ 
  \left\{ 
   - 4 N_S\,\overline{Q^2}\,\left(K_1 + K_2\right) 
   - 4 \left(q_i^2 + q_j^2\right) \left(K_5 + K_6\right) 
  \right.  
 \right. \nonumber \\ 
 && \qquad \quad 
  \left. 
   - 4\,q_i q_j \left(2 K_{18} + K_{19}\right) 
   - 4 N_S\,\overline{Q} \left(q_i + q_j\right) K_{24} 
  \right\} \chi_{ij} \nonumber \\ 
 && \qquad 
  + 2 N_S\,\overline{Q^2} \left(\chi_i + \chi_j\right) K_7  
  \nonumber \\ 
 && \qquad 
  + 2 N_S 
    \left\{ 
     \overline{Q^2} \left(\chi_i + \chi_j\right) 
     + 2\,\overline{\chi}_S\,\left(q_i - q_j\right)^2  
    \right\} K_8 \nonumber \\ 
 && \qquad 
  + 4 \left(q_i^2 \chi_i + q_j^2 \chi_j\right) K_9 
  \nonumber \\ 
 && \qquad 
  + 4 \left\{ 
       q_i^2 \chi_i + q_j^2 \chi_j 
       + \left(q_i - q_j\right)^2 \left(\chi_i + \chi_j\right) 
      \right\} K_{10} \nonumber \\ 
 && \qquad 
 \left. 
  - 4 \left(q_i - q_j\right)^2 \left(\chi_i + \chi_j\right) 
    K_{11} 
  + 4 N_S\, 
    \overline{Q} \left(q_i \chi_i + q_j \chi_j\right) K_{25} 
 \right] 
  \nonumber \\ 
 && 
 + e^4 F_0^2 
   \left[ 
    4 N_S\,\overline{Q^2} \left(q_i - q_j\right)^2 
    \left(K_{15} + K_{16}\right) 
    + 2 (q_i^2 - q_j^2)^2 K_{20} 
   \right. \nonumber \\ 
 && \qquad \quad 
    + 4 (q_i - q_j)^2 \left(q_i^2 + q_j^2\right) K_{21} 
    \nonumber \\ 
 && \qquad \quad 
   \left.  
    + 4 N_S\,\overline{Q} (q_i - q_j)^2 \left(q_i + q_j\right) 
      K_{23}  
   \right] \, . \label{eq:LECcontribution}
\end{eqnarray} 
 In the above, 
\begin{eqnarray} 
 & 
 \displaystyle{
  \overline{\chi}_S 
  \equiv \frac{1}{N_S} \sum_{r=1}^{N_S} \chi_{(r)}\, , 
 } \nonumber \\ 
 & 
 \displaystyle{ 
  \overline{Q} 
  \equiv 
  \frac{1}{N_S} \sum_{r=1}^{N_S} q_{(r)} \, , \quad 
  \overline{Q^2} 
  \equiv 
  \frac{1}{N_S} \sum_{r=1}^{N_S} q_{(r)}^2 \, , 
 }& 
\end{eqnarray} 
with use of electric sea quark charges $q_{(r)} = q_{N_V + r}$. 
 The formulas derived thus far applies 
both to $N_S = 3$ and $N_S = 2$ 
if the sum over sea flavors and 
the one over sea mesons are appropriately understood. 
 The expressions for 
$\left.m^2_{{\rm QCD},\,ij}\right|_4(\infty)$ 
and $\left.m^2_{{\rm EM},\,ij}\right|_4(\infty)$ 
will reduce to the result 
found in Ref.~\cite{Bijnens:2006mk} 
if one sets $N_S = 3$ and $\overline{Q} = 0$ 
and discards the terms of order $e^4$. 

 Before turning to the numerical analysis, 
we observe a few features that can be read off 
from Eq.~(\ref{eq:finiteSizeCorrection}). 
 First, the asymptotic behavior for 
$L \sqrt{\chi_{ij}} \gg 1$ is determined by 
the term including the function $\mathcal{H}(x)$. 
 This term behaves like $1/L$ times the function 
$\mathcal{H}(\sqrt{\chi_{ij}} L)$. 
 As can be seen from Fig.~\ref{fig:functionH}, 
$\mathcal{H}(x)$ gradually increases 
for $x \rightarrow \infty$. 
 We find that the decrease of the finite size scaling effect 
is slightly slower than $1/L$. 

 Secondly, the terms in Eq.~(\ref{eq:finiteSizeCorrection}) 
all vanish for $\sqrt{\chi_{IJ}} \rightarrow 0$ 
except for the one proportional to 
the constant $\kappa$ defined in Eq.~(\ref{eq:defOfkappa}) 
\footnote{ 
 Here we use the fact that 
$\frac{\mathcal{H}(x)}{x}$ is finite for all $x$. 
 This point, however, has been checked 
only through our numerical investigation 
which gives Eq.~(\ref{eq:behaviorOfHForXzero}).}. 
 This term seems to appear 
whatever is used as the low effective field theories,  
because the contribution leading to this term 
originates from the diagrams in scalar QED theory. 
 For instance, that contribution is also involved 
in the model including the vector and axial-vector mesons  
that leads Eq.~(\ref{eq:naiveDiscretizationWithFiniteT}), 
as is suggested from the overall numerical factor. 
 The presence of this term 
indicates that $1/L$ should be regarded as being 
the same order as 
the pseudo-Goldstone mass and the elementary charge $e$ 
in order for the chiral perturbation to remain systematic. 
 In fact, the calculation done thus far 
implicitly assumes the relative magnitude 
of pseudo-Goldstone mass 
and $L$ in the $p$-regime \cite{Colangelo:2004sc}, 
$m_\pi \sim \frac{1}{L} \sim p$.  
 Our result suggests that there is a $p$-regime 
for the systematics chiral perturbation theory 
including electromagnetism in finite volume. 
 The issue examining whether this is actually true 
is beyond the scope of this paper. 

\FIGURE[p]{ 
\caption{profile of function $\mathcal{H}(x)$} 
\epsfig{file=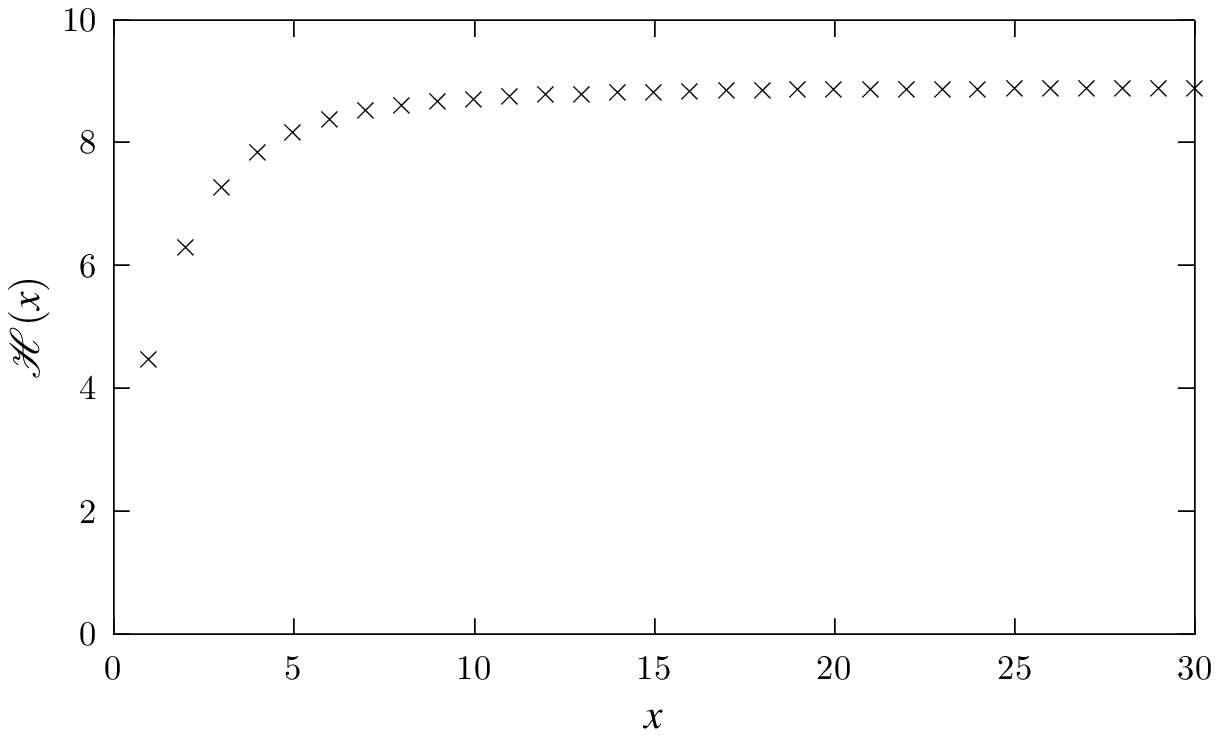,width=14cm}
\label{fig:functionH}
} 

\subsection{numerical investigation}
 We turn to the numerical evaluation of 
the EM correction in the next-to-leading order approximation  
\begin{eqnarray} 
 m^2_{{\rm EM},\,ij}(L) &=& 
 \frac{2 e^2 C}{F_0^2} \left(q_i - q_j\right)^2 
 + \left.m^2_{{\rm EM},\,ij}\right|_4(L) \, , 
  \label{eq:fullEMcorrection}
\end{eqnarray}
to study the qualitative features of 
the QED finite size scaling effect. 
 For that purpose the values of various low energy constants 
found in Ref.~\cite{Bijnens:1996kk} are used as reference 
\begin{eqnarray} 
 && F_0 = 87.7\,{\rm MeV}, \ 
    C = 4.2 \cdot 10^{-5}\,{\rm GeV}^4, \nonumber \\ 
 && L_4 = 0,\ L_5 = 0.97 \cdot 10^{-3},\ 
    L_6 = 0,\ L_8 =0.60 \cdot 10^{-3}, \nonumber \\ 
 && K_5 = 2.85 \cdot 10^{-3}, \ K_9 = 1.3 \cdot 10^{-3}, \ 
 K_{10} = 4.0 \cdot 10^{-3}, \nonumber \\ 
 && K_{11} = -1.25 \cdot 10^{-3} \, . 
\end{eqnarray} 
 The others are set to zero. 
 We employ the formula for $N_S = 3$ and $N_V = 3$ 
and set 
\begin{eqnarray} 
 && 
 \chi_{3} = \chi_{6} = \chi_{9} 
 = (500\ {\rm MeV})^2 \, , \nonumber \\ 
 && 
 q_1 = q_4 = q_7 = \frac{2}{3},\,\quad 
 q_2 = q_3 = q_5 = q_6 = q_8 = q_9 = -\frac{1}{3} \, , 
\end{eqnarray} 
throughout the analysis. 
 In what follows, 
it is always understood that 
the value of the mass of a ghost quark  
is set equal to 
that of the valence quark with the same flavor.

 We plot the dependence of 
$m^2_{{\rm EM},\,12}(L) /m^2_{{\rm EM},\,12}(\infty)$ 
on $L$ in Fig.~\ref{fig:EMchargedPion} 
for the values of quark masses 
corresponding to 
$\chi_1 = \chi_2 = \chi_4 = \chi_5 = (150\ {\rm MeV})^2$ 
and $(300\ {\rm MeV})^2$. 
 The horizontal axis denotes the linear size of volume 
$L$ normalized in unit of $1/M_\rho \simeq 1/(770\ {\rm MeV})$. 
 For instance, for the size 
$L = 16\,a \sim 0.72 \times 1/M_\rho$ 
used in Ref.~\cite{Blum:2007cy}, 
$m^2_{{\rm EM},\,12}(L) /m^2_{{\rm EM},\,12}(\infty)$ 
are 0.144 and 0.421 
for $\chi_1 = (150\ {\rm MeV})^2$ 
and $\chi_1 = (300\ {\rm MeV})^2$, respectively. 
 Thus, our calculation indicates that 
the finite size effect is significant 
for an available lattice geometry. 
 The difference between two quark masses emerges 
for small $L$, and the finite size effect is smaller 
for larger quark mass. 

\FIGURE[p]{ 
\caption{Linear volume size ($L$) dependence 
of the electromagnetic correction in the charged pion mass squared} 
\epsfig{file=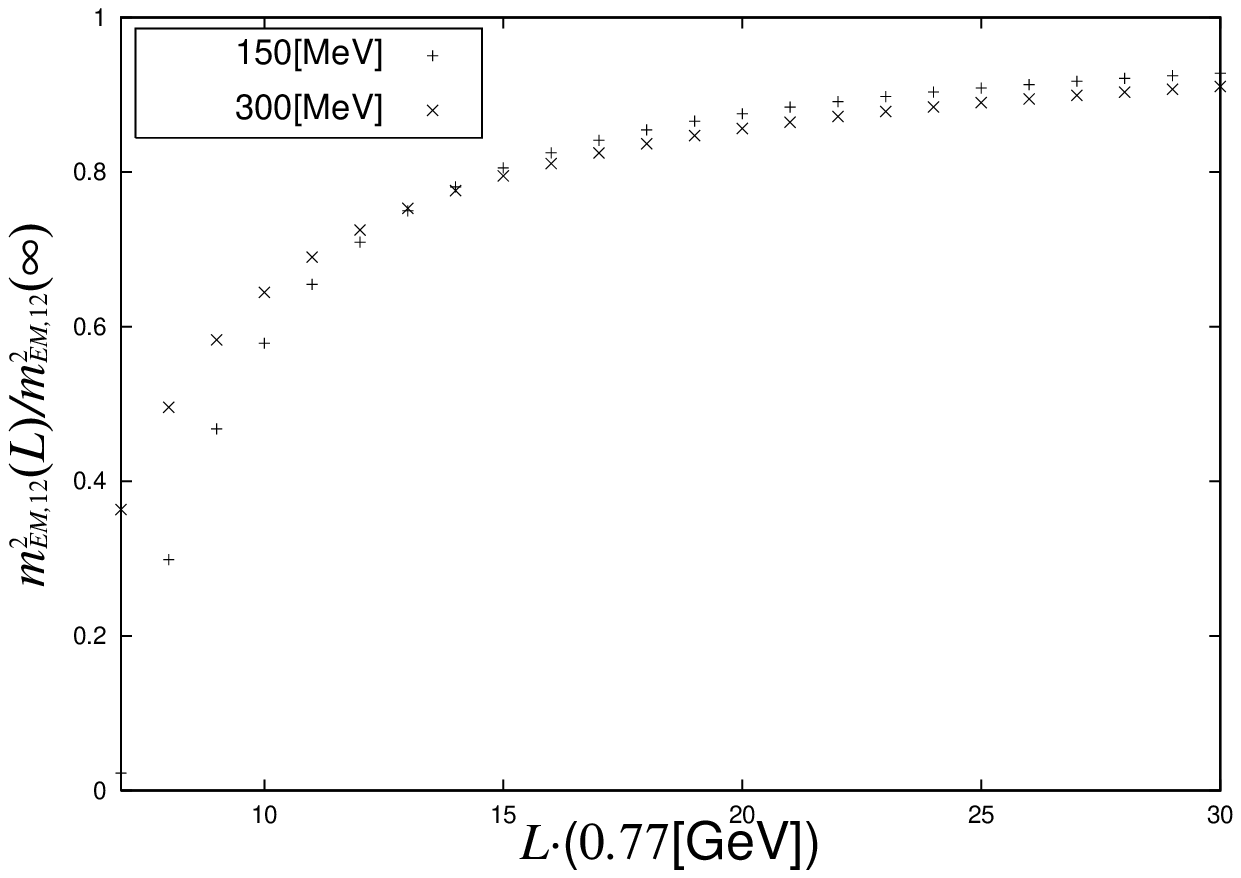,width=14cm}
\label{fig:EMchargedPion}
}

\FIGURE[p]{ 
\caption{Linear volume size ($L$) dependence 
of the electromagnetic splitting ($\Delta m_K^2$) 
in the kaon mass squared} 
\epsfig{file=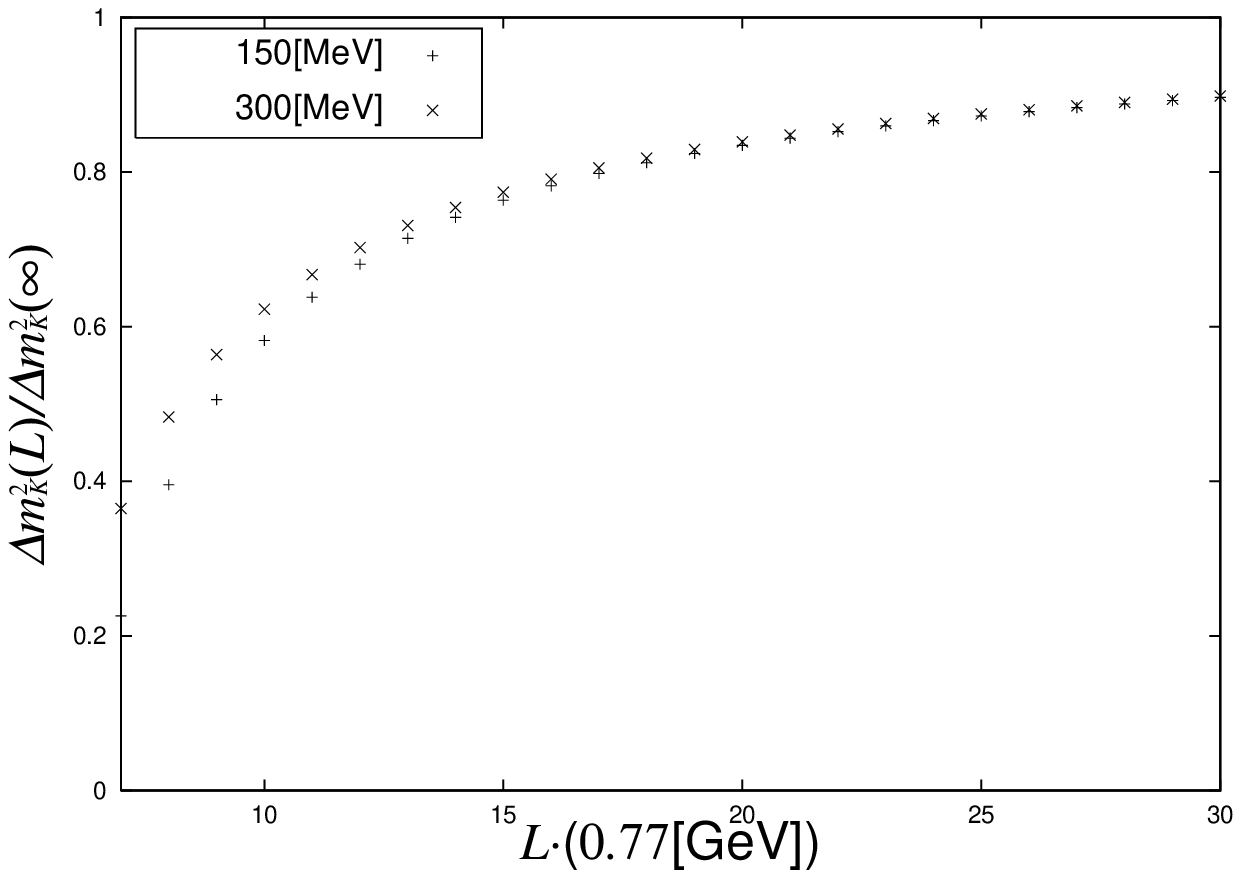,width=14cm} 
\label{fig:EMsplitKaon}
}
 
 We next study the EM splitting $\Delta m^2_{{\rm K}}(L)$ 
in the (valence) kaon mass squared. 
 Using $m^2_{{\rm EM},\,ij}(L)$ in Eq.~(\ref{eq:fullEMcorrection}),  
it is given by  
\begin{eqnarray} 
 \Delta m^2_{{\rm K}}(L) 
 &\equiv& 
 m^2_{{\rm EM},\,13}(L)  
 - m^2_{{\rm EM},\,23}(L) \, .  
\end{eqnarray} 
 Figure \ref{fig:EMsplitKaon} 
shows the $L$-dependence 
of $\Delta m^2_{{\rm K}}(L) /\Delta m^2_{{\rm K}}(\infty)$ 
for the two sets of quark masses corresponding to 
the same values 
of $\chi_1 = \chi_2 = \chi_4 = \chi_5$ 
used in Fig.~\ref{fig:EMchargedPion} 
with $\chi_3 = (500\ {\rm MeV})^2$ fixed. 
 The $L$-dependence in Fig.~\ref{fig:EMsplitKaon} 
is almost similar to that in Fig.~\ref{fig:EMchargedPion} 
for each set of masses. 
 For small $L$ and $\chi_1 = (150\ {\rm MeV})^2$, 
the size of finite size scaling effect is smaller 
than the electromagnetic correction to the charged pion mass. 
 For instance, for $L = 16\,a$, 
$\Delta m^2_{{\rm K}}(L) /\Delta m^2_{{\rm K}}(\infty)$ 
is $0.299$ for $\chi_1 = (150\ {\rm MeV})^2$ 
and $0.415$ for $\chi_1 = (500\ {\rm MeV})^2$. 
 Figures \ref{fig:EMchargedPion} and \ref{fig:EMsplitKaon} 
show that the values of EM correction 
depends on the quark masses for $L = 16\,a$. 
 This observation indicates that 
the sizes of the terms depending on the quark masses 
in Eq.~(\ref{eq:finiteSizeCorrection})
are comparable to 
or more important than 
that of the term proportional to $\kappa$ 
for the quark masses used in the present analysis. 

\FIGURE[p]{ 
\caption{Finite size scaling effect 
on the EM splitting ($\Delta m_K^2$) in kaon mass squared 
in partially quenched QCD, 
where open squares stand for  
the $L$-dependence for 
$\chi_1 = \chi_2 = 150\ {\rm MeV}$, 
$\chi_4 = \chi_5 = 300\ {\rm MeV}$,  
and dark squares with cross marks stand for 
the one for $\chi_1 = \chi_2 = 300\ {\rm MeV}$, 
$\chi_4 = \chi_5 = 150\ {\rm MeV}$. 
 No significant change is observed between them. 
 They differ from the other plots corresponding 
to unquenched QCD in the small volume region. 
} 
\epsfig{file=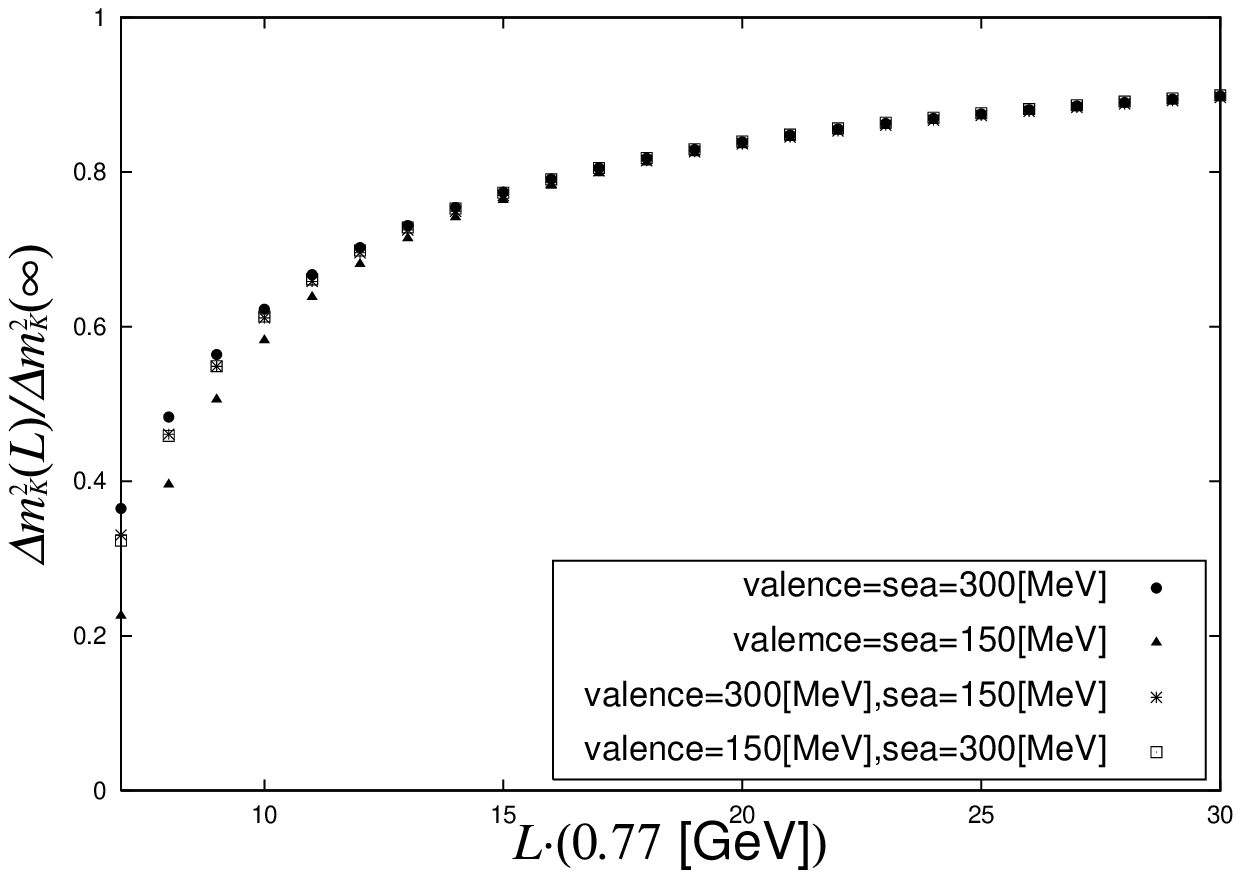,width=14cm}
\label{fig:PartiallyQuenched}
}

 Figure \ref{fig:PartiallyQuenched} 
compares the $L$-dependence 
of the electromagnetic splitting of $m_K^2(L)$ 
in partially quenched QCD with that in (unquenched) QCD. 
 The two sets of plots are drawn 
in Fig.~\ref{fig:PartiallyQuenched} for 
\begin{eqnarray} 
 {\rm set\ 1}\ &\Leftrightarrow&\ 
 \chi_1 = \chi_2 = (150\ {\rm MeV})^2,\  
 \chi_4 = \chi_5 = (300\ {\rm MeV})^2, \nonumber \\ 
 {\rm set\ 2} &\Leftrightarrow&  
 \chi_1 = \chi_2 = (300\ {\rm MeV})^2,\  
 \chi_4 = \chi_5 = (150\ {\rm MeV})^2 \, . 
\end{eqnarray} 
 As can be seen, these two sets of plots 
in partially quenched QCD coincide with each other. 
 By evaluating the $L$-dependence for various set of the values  
of quark masses, we find that 
the $L$-dependence for a set of the values of quark masses 
is almost the same as that for the set with 
the valence and sea quark masses interchanged. 
 This observation together with the comparison of 
the relative sizes of four sets of plots 
in Fig.~\ref{fig:PartiallyQuenched} 
shows that 
the size of the finite size correction 
is roughly determined by the average value of the quark masses 
involved irrespective of whether 
the system is unquenched or partially-quenched. 
 To elucidate this point, we show 
in Figs.~\ref{fig:MassDependence_16} 
and \ref{fig:MassDependence_32} 
the quark mass dependence 
of the finite size scaling effect on $m_K^2(L)$ 
on $m^2 \equiv \chi_1 = \chi_2 = 2 B_0 m_u$ 
for fixed $L = 16\,a$ and $L = 32\,a$ 
($a \simeq 1/(1.66\ {\rm GeV})$ \cite{Blum:2007cy}), 
respectively. 
 In the partially quenched case, 
the masses of up and down sea quarks 
are fixed to be $300\ {\rm MeV}$. 
 We can see that, at $L = 16\,a$,  
the size of the finite size effect changes rapidly 
for $m^2 \lesssim 0.05\ {\rm [GeV]^2}$ in the unquenched case, 
while no such significant change is observed  
in the partially quenched case. 
 For larger $m^2$, they approach to each other. 
 For $L = 32\,a$ and larger $L$, 
the finite size effect is determined by 
the relative size of $L$ 
and $\chi_3 = \chi_6 = (500\ {\rm MeV})^2$.

\FIGURE[p]{ 
\caption{Dependence of the finite size scaling effect 
on the electromagnetic splitting ($\Delta m_K^2$) 
in kaon mass squared 
on $m^2 = \chi_1$ at the tree level 
for $L = 16\,a$. 
 The dark circular dots represent the unquenched case 
while the inverted triangles represent 
the valence quark mass dependence in partially quenched QCD. 
} 
\epsfig{file=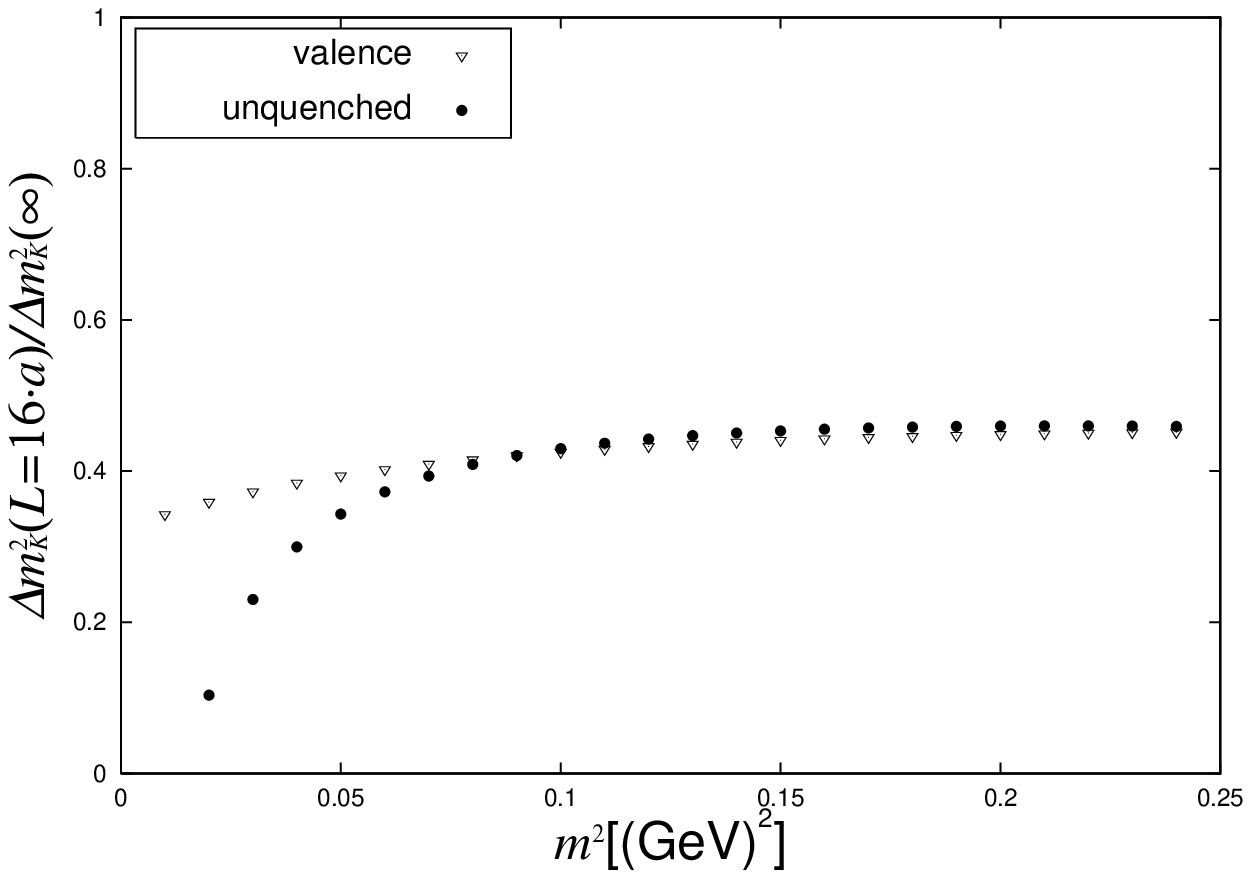,width=14cm}
\label{fig:MassDependence_16}
}

\FIGURE[p]{ 
\caption{Dependence of the finite size scaling effect 
on the electromagnetic splitting 
($\Delta m_K^2$) in kaon mass squared 
on $m^2 = \chi_1$ at the tree level 
for $L = 32\,a$. 
} 
\epsfig{file=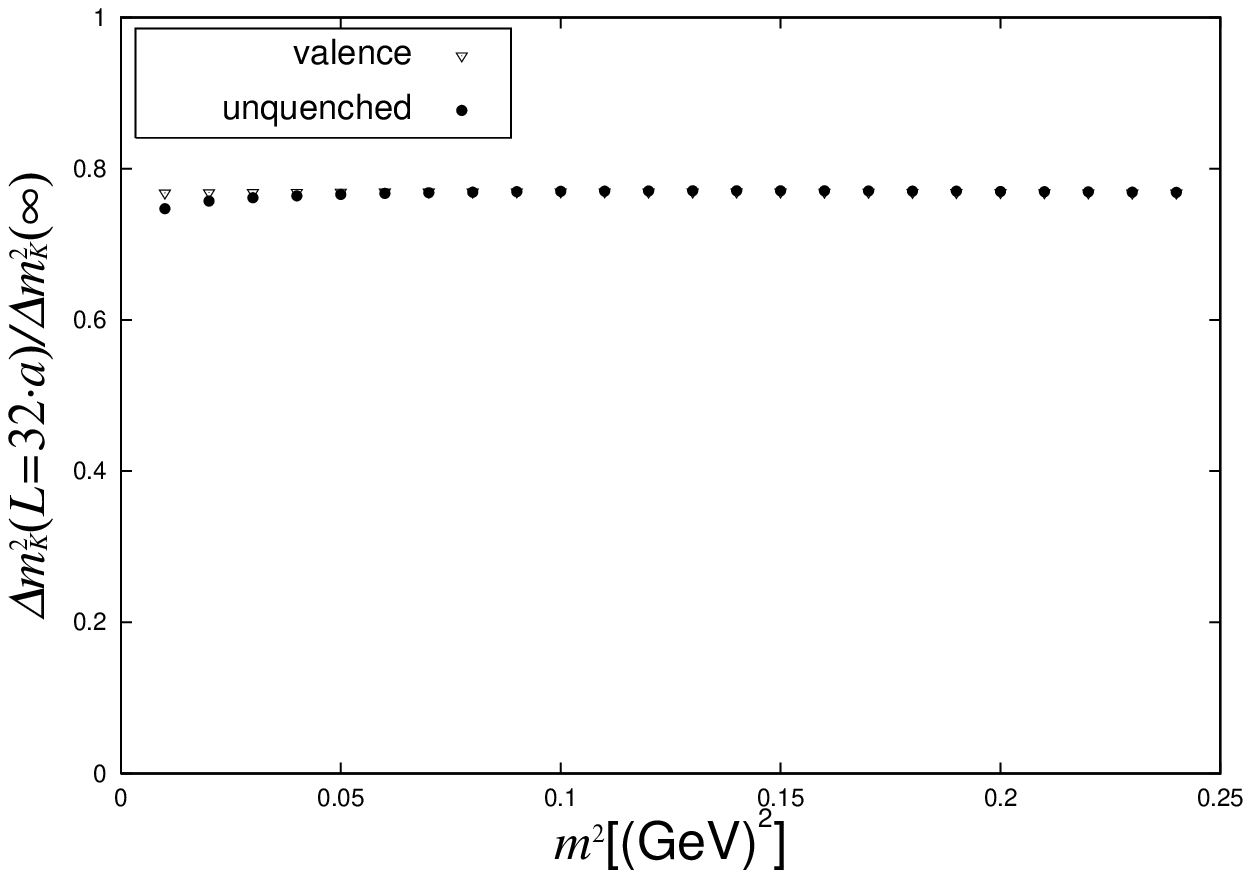,width=14cm}
\label{fig:MassDependence_32}
}

\section{Conclusion and discussion} 
\label{sec:discussion} 

 In this paper, we studied the finite size scaling effect 
on the electromagnetic (EM) correction to the pseudoscalar meson 
masses from the low energy effective field theory 
of QCD including electromagnetism. 
 For that purpose, we began with constructing a new QED 
in finite volume, QED$_{L}$.
 Taking the practical application to the lattice simulation 
into account, 
we adapted it to the partially quenched chiral perturbation theory
including electromagnetism. 
 We computed the electromagnetic correction to
the pseudoscalar meson mass squared at the next-to-leading order
on both of the spaces with finite and infinite volumes 
for generic number $N_S$ of sea quarks. 
 Through numerical investigation for $N_S = 3$, 
we found that the finite size scaling effect on the EM correction
is sizable on the space with the volume available 
in the lattice simulation.
 By investigating its dependence on the quark masses 
in unquenched and partially quenched systems, 
we pointed out that the finite size correction
is determined by the averaged values of masses of quarks 
involved in the system. 
 Though the current study was restricted to 
the pseudoscalar meson masses, we can study the EM corrections 
to the other hadronic observables such as decay constants 
in finite volume. 

 It should be noted that the finite size correction 
tends to increase for the increase of the extent $T$ 
of the temporal direction. 
 For example, let us suppose that 
the value $\sim 0.9$ in Eq.~(\ref{eq:RBCvalue}) 
does not change so drastically 
even if the contribution 
of $\widetilde{A}_\mu(t,\,{\bf 0})$ is subtracted. 
 The explicit calculation 
in the low energy effective theory 
including vector and axial-vector mesons 
show that it is reduced to about $0.6$ 
in the limit $T \rightarrow \infty$. 
 Since the simulation has to be performed with finite $T$, 
$m^2_{{\rm EM},\,12}(L) /m^2_{{\rm EM},\,12}(\infty)$ 
obtained in our present analysis with $T \rightarrow \infty$ 
may overestimate the finite size correction 
in the actual simulation. 
 It is one of the subjects to fill the gap 
between these two.  

 Finally, we recall that we construct 
QED$_{L}$ 
because we would like to respect the boundary condition 
(i.e., periodic boundary condition)
for the meson fields along spatial directions 
usually assumed in the lattice QCD simulations, 
and to understand the qualitative properties 
of the finite size scaling effect 
with the practical situation in our mind. 
 It is plausible that the finite size scaling effect 
depends on the boundary conditions imposed on fields. 
 As illustrated in Figs.~\ref{fig:PartiallyQuenched}, 
if periodic boundary condition is employed, 
the dependence on the volume is not simple 
around the available volume 
$L \cdot (0.77\ {\rm GeV}) \lesssim 10$ in the lattice simulations. 
 It would be one of the important subjects 
to study the finite size scaling effect on EM corrections 
for the various types of boundary conditions 
and to select the one 
in which the finite size effect exhibits such a simplest behavior 
that allows us to extrapolate the data at available volumes 
to the value in infinite volume.

\acknowledgments{ 
 We thank T.~Blum, T.~Doi, T.~Izubuchi, 
N.~Yamada, K.~Yamawaki and R.~Zhou 
for valuable discussions. 
 This work is supported in part by JSPS Grant-in-Aid 
for Scientific Research (C) 20540261 
and MEXT Grant-in-Aid 
for Scientific Research on Priority Areas 
``Higgs and Supersymmetry'' 20025010. 

}

\appendix 

\section{Formulae for several sums} 
\label{sec:basicSums}
 
 The one-loop corrections to meson masses 
in chiral perturbation theory 
for QCD plus QED system with finite volume 
are described in terms of several basic functions. 
 Each of these functions takes the form 
of the one-dimensional integral 
of the sum over three-dimensional momenta 
of some function. 
 We employ dimensional regularization 
and define 
\begin{eqnarray} 
 & 
 \displaystyle{ 
  \int_{-\infty}^\infty \frac{dk^0}{2\pi i}\, 
  \frac{1}{V} \sum_{{\bf k} \in \widetilde{\Gamma}_d} 
  = 0\, . 
 }& 
\end{eqnarray}  
 Here, $d \equiv D - 1$, 
$V \equiv L^d$ and the sum runs over all 
${\bf k} \in \widetilde{\Gamma}_d$, where 
\begin{eqnarray} 
 & 
 \displaystyle{ 
  \widetilde{\Gamma}_d \equiv 
  \left\{ 
   {\bf k} = \left(k^1,\,\cdots,\,k^d\right)\, 
   \left|\, 
    k^j \in \frac{2\pi}{L}\,\mathbb{Z} 
   \right. 
  \right\} \, . 
 }& \label{eq:def_tildeGamma_d}
\end{eqnarray} 
 The aim of this Appendix 
is to get compact expressions for the four functions 
that can be evaluated by {\sf MATHEMATICA} and so forth 
by following the strategy in Ref.~\cite{AliKhan:2003cu}. 

 We first consider the function 
\begin{eqnarray} 
 & 
 \displaystyle{ 
  I_{11}(m^2;\,L) 
  \equiv 
  \left(\mu^2\right)^{2 - \frac{D}{2}} 
  \int_{-\infty}^\infty \frac{dk^0}{2\pi i} 
  \frac{1}{V}
  \sum_{{\bf k} \in \widetilde{\Gamma}_d^\prime}  
  \frac{1}{\left(-k^2 - i \epsilon\right) 
           \left\{m^2 - \left(k+p\right)^2 - i\epsilon\right\}} \, . 
 }& \label{eq:sum_I11}
\end{eqnarray} 
 In the above, $p$ is assumed to be on-shell, $p^2 = m^2$ 
and 
$\widetilde{\Gamma}_d^\prime \equiv 
\widetilde{\Gamma}_d - \left\{{\bf 0}\right\}$. 
 
 By introducing the Feynman parameters as usual, 
Eq.~(\ref{eq:sum_I11}) becomes 
\begin{eqnarray} 
 & 
 \displaystyle{ 
  I_{11}(m^2;\,L) 
  \equiv 
  \int_0^1 dy 
  \int_{-\infty}^\infty \frac{d k^0}{2\pi i} 
  \frac{1}{V} \sum_{{\bf k} \in \widetilde{\Gamma}_d^\prime} 
  \frac{1}{\left\{y^2 m^2 - (k+yp)^2 - i \epsilon\right\}^2} \, , 
 }& \label{eq:I11WithFeynmanParameters}
\end{eqnarray} 
 The sum of a function $\widetilde{F}({\bf k})$ over 
${\bf k} \in \widetilde{\Gamma}_d^\prime$ can be written 
as an integral 
\begin{eqnarray} 
 & 
 \displaystyle{ 
  \frac{1}{V} \sum_{{\bf k} \in \widetilde{\Gamma}_d^\prime} 
   \widetilde{F}({\bf k}) 
  = 
  \int d^d k^\prime \widetilde{F}({\bf k}^\prime)\, 
  \frac{1}{V} \sum_{{\bf k} \in \widetilde{\Gamma}_d} 
   \delta^d\left({\bf k}^\prime - {\bf k}\right)    
  - 
  \frac{1}{V}\,\widetilde{F}({\bf 0}) \, . 
 }& \label{eq:sumOfF}
\end{eqnarray} 
 Using the Poisson resummation 
\begin{eqnarray} 
 & 
 \displaystyle{ 
  \frac{1}{V} \sum_{{\bf k} \in \widetilde{\Gamma}_d} 
  \delta^d\left({\bf k}^\prime - {\bf k}\right)  
  = 
  \frac{1}{\left(2\pi\right)^d} 
  \sum_{{\bf x} \in \Gamma_d} e^{i {\bf k} \cdot {\bf x}} \, , 
 }& 
\end{eqnarray} 
where 
\begin{eqnarray} 
 & 
 \displaystyle{ 
  \Gamma_d \equiv 
  \left\{ 
   {\bf x} \equiv \left(x^1,\,\cdots,\,x^d\right)\, 
   \left| 
    x^j \in L \mathbb{Z} 
   \right. 
  \right\} \, , 
 }& 
\end{eqnarray} 
the sum (\ref{eq:sumOfF}) can be written as 
\begin{eqnarray} 
 & 
 \displaystyle{ 
  \frac{1}{V} \sum_{{\bf k} \in \widetilde{\Gamma}_d^\prime} 
   \widetilde{F}({\bf k}) 
  = 
  \sum_{{\bf x} \in \Gamma_d} F({\bf x}) 
  - 
  \frac{1}{V}\,\int d^d x\,F({\bf x}) \, , 
 }& 
\end{eqnarray} 
where $F({\bf x})$ 
is obtained by Fourier transformation of $\widetilde{F}({\bf k})$ 
\begin{eqnarray} 
 & 
 \displaystyle{ 
  F({\bf x}) = 
  \int \frac{d^d k}{\left(2\pi\right)^d}\, 
   e^{i {\bf k} \cdot {\bf x}}\,\widetilde{F}({\bf k}) \, . 
 }& \label{eq:FourierTransform}
\end{eqnarray} 
 In the present context $\widetilde{F}({\bf k})$ is 
\begin{eqnarray}  
 \widetilde{F}({\bf k}) &=& 
 \frac{1} 
      {\left\{y^2 m^2 - \left(k + yp\right)^2 - i \epsilon\right\}^2} 
  \nonumber \\ 
 &=& 
 \frac{1}{\Gamma(2)} 
 \int_0^\infty \frac{d\lambda}{\lambda} \left(i \lambda\right)^2 
 \exp 
 \left[ 
  - i \lambda 
  \left\{ 
   y^2 m^2 - \left(k + zp\right)^2 - i \epsilon
  \right\} 
 \right] \, . \label{eq:tildedFOfOurInterest}
\end{eqnarray} 

 A given $d$-dimensional vector ${\bf x}$
defines a $D$-dimensional space-like vector 
$x^\mu = (0,\,{\bf x})$.  
 A one-dimensional integral in Eq.~(\ref{eq:I11WithFeynmanParameters}) 
and a $d$-dimension integral in Eq.~(\ref{eq:FourierTransform}) 
combine to become an integral over $D$-dimensional momenta, 
which can be readily carried out;  
\begin{eqnarray} 
 && 
 \int_0^\infty \frac{d\lambda}{\lambda} \left(i\lambda\right)^2 
 \int 
 \frac{d^D k}{i \left(2\pi\right)^D} 
 \exp 
 \left[ 
  - i \lambda 
  \left\{ 
   - \left(k + yp\right)^2 - i \epsilon
  \right\} 
  - i x \cdot k 
 \right] \nonumber \\ 
 && \quad 
 = 
 \int_0^\infty \frac{d\lambda}{\lambda}
 \frac{\lambda^2}{\left(4\pi\lambda\right)^{\frac{D}{2}}} 
 \exp 
 \left[ 
  - \lambda 
  \left\{ 
   \left(i \frac{x}{2\lambda} - yp\right)^2 - (yp)^2 
  \right\} 
 \right] \, . 
\end{eqnarray} 
 For $p^\mu = (m,\,{\bf 0})$, 
$\left(i \frac{x}{2\lambda} - yp\right)^2 - (yp)^2 = 
\frac{\left|{\bf x}\right|^2}{4 \lambda^2}$. 
 Eq.~(\ref{eq:I11WithFeynmanParameters}) thus becomes 
\begin{eqnarray} 
 I_{11}(m^2;\,L) 
 &=& 
 \left(\mu^2\right)^{2 - \frac{D}{2}}
 \int_0^1 dy 
 \left( 
  \sum_{{\bf n} \in \mathbb{Z}^d} 
  - \int d^d n 
 \right) 
  \nonumber \\ 
 && \quad 
 \times 
 \int_0^\infty \frac{d\lambda}{\lambda}\, 
 \frac{\lambda^2}{\left(4\pi\lambda\right)^{\frac{D}{2}}}\, 
 \exp 
 \left( 
  -\lambda\,y^2 m^2 - \frac{L^2}{4 \lambda} \left|{\bf n}\right|^2  
 \right) \, . 
\end{eqnarray} 
 In the sum appearing above, 
the contribution of ${\bf n} = {\bf 0}$ is exactly 
$I_{11}(m^2) = I_{11}(m^2;\,L \rightarrow \infty)$. 
 $I_{11}(m^2;\,L) - I_{11}(m^2)$ is hence free of UV divergence. 
 Letting $D \rightarrow 4$ for this difference 
and rescaling $\lambda \rightarrow \frac{L^2}{4\pi} \lambda$ 
leads 
\begin{eqnarray} 
 I_{11}(m^2;\,L) - I_{11}(m^2) 
 = 
 - 
 \frac{1}{16 \pi^2}\,\frac{1}{mL}\, 
 \mathcal{H}(mL) \, . 
\end{eqnarray} 
 Here  
\begin{eqnarray} 
 \mathcal{H}(mL) 
 &\equiv& 
 \pi 
 \int_0^\infty \frac{d\lambda}{\lambda^{\frac{3}{2}}}\, 
 {\rm erf}\left(mL \sqrt{\frac{\lambda}{4\pi}}\right) 
 \mathcal{S}(\lambda) \, , \label{eq:fun_H} \\ 
 \mathcal{S}(\lambda) 
 &\equiv&   
 - 
 \left( 
  \sum_{{\bf n} \in \mathcal{Z}^3 - \left\{{\bf 0}\right\}} 
  - \int d^3 n  
 \right)  
 \exp\left(- \frac{\pi}{\lambda} \left|{\bf n}\right|^2\right) 
  \nonumber \\ 
 &=& 
 - 
 \left\{ 
  \left(\vartheta_3\left(0,\,i \frac{1}{\lambda}\right)\right)^3 
  - 1 - \lambda^{\frac{3}{2}} 
 \right\} \, , \label{eq:fun_S}
\end{eqnarray} 
where ${\rm erf}(x)$ is the error function 
\begin{eqnarray} 
 & 
 \displaystyle{ 
  {\rm erf}(x) = \frac{2}{\sqrt{\pi}} \int_0^x ds\,e^{-s^2} 
  \, , 
 }& 
\end{eqnarray} 
and $\vartheta_3(v;\,\tau)$ is a Jacobi-theta function 
\begin{eqnarray} 
 \vartheta_3(v;\,\tau) 
 = 
 \sum_{n=-\infty}^\infty 
 \exp\left(\pi \tau i n^2 + 2\pi v i n\right) \, . 
\end{eqnarray} 
 We recall that 
the term $\lambda^{\frac{3}{2}}$ in $\mathcal{S}(\lambda)$ 
arises in our new QED, QED$_L$. 
 Because 
\begin{eqnarray} 
 \vartheta_3\left(0,\,i\,\frac{1}{\lambda}\right) 
 \rightarrow \lambda^{\frac{1}{2}} \, , 
\end{eqnarray} 
in the infrared limit $\lambda \rightarrow \infty$, 
the presence of that term indeed ensures IR-finiteness of 
the integral over $\lambda$ in the expression (\ref{eq:fun_H}). 
 The numerical study shows that 
\begin{eqnarray} 
 \lim_{x \rightarrow 0} \frac{\mathcal{H}(x)}{x} 
 \simeq 10.4 \, . \label{eq:behaviorOfHForXzero} 
\end{eqnarray} 

 Next we consider the function 
\begin{eqnarray} 
 & 
 \displaystyle{ 
  J_{11}(m^2;\,L) 
  \equiv 
  \left(\mu^2\right)^{2 - \frac{D}{2}} 
  \int_{-\infty}^\infty \frac{dk^0}{2\pi i} 
  \frac{1}{V}
  \sum_{{\bf k} \in \widetilde{\Gamma}_d^\prime}  
  \frac{p \cdot k} 
       {\left(-k^2 - i \epsilon\right) 
        \left\{m^2 - \left(k+p\right)^2 - i\epsilon\right\}} \, . 
 }& \nonumber \\ \label{eq:sum_J11}
\end{eqnarray} 
 The same manipulation as done for $I_{11}(m^2;\,L)$ 
in Eq.~(\ref{eq:tildedFOfOurInterest}) leads 
\begin{eqnarray} 
 J_{11}(m^2;\,L) 
 &=& 
 \int_0^1 dy 
 \left(\sum_{{\bf x} \in \Gamma_d} - \frac{1}{V} \int d^d x\right) 
  \nonumber \\ 
 && \ 
 \times 
 \frac{1}{\Gamma(2)} 
 \int_0^\infty \frac{d\lambda}{\lambda} 
 \left(i\lambda\right)^2 
 \left(\mu^2\right)^{2 - \frac{D}{2}}
 \int \frac{d^D k}{i \left(2\pi\right)^D} \nonumber \\ 
 && \ 
 \times 
 p \cdot k  
 \exp 
 \left[ 
  - (i\lambda) 
    \left\{ 
     y^2 m^2 - \left(k + yp\right)^2 - i \epsilon 
    \right\} 
  - i x \cdot k 
 \right] \, . 
\end{eqnarray} 
 We follow Appendix of Ref.~\cite{Hayakawa:1999zf} 
to carry out such integrals with $k_\mu$ in the numerator. 
 We replace $k_\mu$ appearing in the numerator as 
\begin{eqnarray} 
 & 
 \displaystyle{ 
  k_\mu = 
  \left[ 
   - \frac{1}{2}\,\frac{1}{i \lambda}\,\frac{\del}{\del \rho^\mu}
   \exp\left(- 2 \left(i\lambda\right) \rho \cdot k\right) 
  \right]_{\rho \rightarrow 0} \, . 
 }& 
\end{eqnarray} 
 Carrying out the integral of $k$ and performing 
a $\rho$-derivative gives  
\begin{eqnarray} 
 J_{11}(m^2;\,L) 
 &=& 
 \int_0^1 dy 
 \left(\sum_{{\bf x} \in \Gamma_d} - \frac{1}{V} \int d^d x\right)  
 \int_0^\infty \frac{d\lambda}{\lambda} 
 \frac{\lambda^2}{\left(4 \pi \lambda\right)^{\frac{D}{2}}} 
  \nonumber \\ 
 && \quad 
 \times 
 p \cdot \left(i\,\frac{x}{2 \lambda} - y p\right) 
 \exp 
 \left[ 
  - \lambda \left(i\,\frac{x}{2\lambda} - y p\right)^2 
 \right] \nonumber \\ 
 &=& 
 - \frac{1}{2}
 \left(\sum_{{\bf x} \in \Gamma_d} - \frac{1}{V} \int d^d x\right) 
 \int_0^\infty \frac{d\lambda}{\lambda} 
 \frac{\lambda}{\left(4 \pi \lambda\right)^{\frac{D}{2}}} 
 \left(1 - e^{-\lambda m^2}\right) 
 e^{-\frac{\left|{\bf x}\right|^2}{4 \lambda}} \, , 
  \nonumber \\ 
  \label{eq:J11FV_temp}
\end{eqnarray} 
where the second equality 
follows by taking $p^\mu = (m,\,{\bf 0})$ 
and performing the integral over $y$. 
 In Eq.~(\ref{eq:J11FV_temp}), 
UV divergence is contained in the term with ${\bf x} = {\bf 0}$, 
which is exactly 
$J_{11}(m^2) = J_{11}(m^2,\,L \rightarrow \infty)$. 
 Therefore, $J_{11}(m^2;\,L) - J_{11}(m^2)$ is UV-finite. 
 In terms of the function 
\begin{eqnarray} 
 & 
 \displaystyle{ 
  \mathcal{K}(x) 
  \equiv 
  4\pi \int_0^\infty \frac{d\lambda}{\lambda}\,\frac{1}{\lambda} 
  \left(1 - e^{-\frac{x^2}{4\pi}\,\lambda}\right) 
  \mathcal{S}(\lambda) \, , 
 }& 
\end{eqnarray} 
it can be expressed as 
\begin{eqnarray} 
 J_{11}(m^2;\,L) - J_{11}(m^2) 
 &=& 
 \frac{1}{32\pi^2}\,\frac{1}{L^2}\,\mathcal{K}(mL) \, . 
\end{eqnarray} 

 The quantity 
\begin{eqnarray} 
 I_1^0(L) &\equiv& 
 \left(\mu^2\right)^{2 - \frac{D}{2}}
 \int_{-\infty}^\infty \frac{d k^0}{2\pi i} 
 \frac{1}{V} \sum_{{\bf k} \in \widetilde{\Gamma}_d^\prime}  
 \frac{1}{- k^2 - i \epsilon} \, ,  
  \label{eq:sum_I_1-0}
\end{eqnarray} 
needs a special care. 
 In dimensional regularization, we put 
\begin{eqnarray} 
 & 
 \displaystyle{ 
  I_1^0(\infty) 
  = 
  \left(\mu^2\right)^{2 - \frac{D}{2}}
  \int \frac{d^D k}{i \left(2\pi\right)^D}\, 
   \frac{1}{- k^2 - i \epsilon}  
  = 0 \, . 
 }& \nonumber 
\end{eqnarray} 
 However, we cannot set $I_1^0(L) = 0$.  
 For instance, if the integral 
\begin{eqnarray} 
 && 
 \left(\mu^2\right)^{2 - \frac{D}{2}}
 \int_{-\infty}^\infty \frac{dk^0}{2\pi i} 
 \frac{1}{L^d} 
 \sum_{{\bf k} \in \widetilde{\Gamma}_d^\prime} 
 \frac{1}{\left(-k^2 - i \epsilon\right) 
          \left(m^2 - k^2 - i \epsilon\right)} 
  \nonumber \\ 
 && \qquad 
 =  
 \frac{1}{m^2} 
 \left(\mu^2\right)^{2 - \frac{D}{2}}
 \int_{-\infty}^\infty \frac{dk^0}{2\pi i} 
 \frac{1}{L^d} 
 \sum_{{\bf k} \in \widetilde{\Gamma}_d^\prime} 
 \left( 
  \frac{1}{- k^2 - i \epsilon} 
  - 
  \frac{1}{m^2 - k^2 - i \epsilon} 
 \right) \, , 
\end{eqnarray} 
were evaluated in two different ways; 
(1) direct evaluation of the left-hand side by introducing 
a Feynman parameter, 
and (2) evaluation of the right-hand only with the second term kept, 
inconsistency would arise.   
 A straightforward calculation yields 
\begin{eqnarray} 
 & 
 \displaystyle{ 
  I_1^0(L) - I_1^0(\infty) 
  = 
  - \frac{\kappa}{4\pi}\,\frac{1}{L^2} \, , 
 }&  
\end{eqnarray} 
where $\kappa$ is a constant defined by 
\begin{eqnarray} 
 & 
 \displaystyle{ 
  \kappa \equiv 
  \int_0^\infty \frac{d\lambda}{\lambda^2}\,\mathcal{S}(\lambda) 
  \approxeq 2.837 \, . 
 }& \label{eq:defOfkappa}
\end{eqnarray} 

 All the one-loop contributions induced by quartic couplings 
are described by a function 
\begin{eqnarray} 
 I_1(m^2;\,L) 
 &\equiv& 
 \left(\mu^2\right)^{2 - \frac{D}{2}}
 \int_{-\infty}^\infty \frac{d k^0}{2\pi i} 
 \frac{1}{V} \sum_{{\bf k} \in \widetilde{\Gamma}_d}  
 \frac{1}{m^2 - k^2 - i \epsilon} \, . 
%
 \label{eq:sum_I_1}
\end{eqnarray} 
 The expression for $I_1(m^2;\,L)$ 
in terms of Jacobi-theta function 
was obtained in Ref.~\cite{AliKhan:2003cu} 
\begin{eqnarray} 
 I_1(m^2;\,L) - I_1(m^2) &=& 
 \frac{1}{\left(4\pi\right)^2}\,\frac{\mathcal{M}(mL)}{L^2} 
  \, , 
\end{eqnarray} 
where 
\begin{eqnarray} 
 \mathcal{M}(x) 
 &\equiv& 
 4\pi \int_0^\infty \frac{d\lambda}{\lambda^2}\, 
 \exp\left(-\frac{x^2}{4\pi}\,\lambda\right) 
 \mathcal{T}(\lambda) \, , 
\end{eqnarray} 
and 
\begin{eqnarray} 
 \mathcal{T}(\lambda) 
 &\equiv& 
 \sum_{{\bf n} \in 
       \left(\mathbb{Z}^3 - \left\{{\bf 0}\right\}\right)} 
 \exp\left(-\frac{\pi}{\lambda}\,\left|{\bf n}\right|^2\right) 
  \nonumber \\ 
 &=& 
 \left(\theta_3\left(0,\,i\,\frac{1}{\lambda}\right)\right)^3 
 - 1 \, . 
\end{eqnarray} 
\bibliographystyle{JHEP} 
\bibliography{latticeMassLEC,finiteVolume,chpt,etcRef} 

\providecommand{\href}[2]{#2}\begingroup\raggedright\begin{thebibliography}{10}

\bibitem{AliKhan:2001tx}
{\bf CP-PACS} Collaboration, A.~Ali~Khan {\em et.~al.}, {\it {\it Light hadron
  spectroscopy with two flavors of dynamical quarks on the lattice}},  {\em
  Phys. Rev.} {\bf D {\bf 65}} (2002) 054505,
  [\href{http://xxx.lanl.gov/abs/hep-lat/0105015}{{\tt hep-lat/0105015}}].

\bibitem{Aoki:2002uc}
{\bf JLQCD} Collaboration, S.~Aoki {\em et.~al.}, {\it {\it Light hadron
  spectroscopy with two flavors of $O(a)$-improved dynamical quarks}},  {\em
  Phys. Rev.} {\bf D {\bf 68}} (2003) 054502,
  [\href{http://xxx.lanl.gov/abs/hep-lat/0212039}{{\tt hep-lat/0212039}}].

\bibitem{Gockeler:2004rp}
{\bf QCDSF} Collaboration, M.~Gockeler {\em et.~al.}, {\it {\it Determination
  of light and strange quark masses from full lattice QCD}},  {\em Phys. Lett.}
  {\bf {\bf B639}} (2006) 307--311,
  [\href{http://xxx.lanl.gov/abs/hep-ph/0409312}{{\tt hep-ph/0409312}}].

\bibitem{DellaMorte:2005kg}
{\bf ALPHA} Collaboration, M.~Della~Morte {\em et.~al.}, {\it {\it
  Non-perturbative quark mass renormalization in two-flavor QCD}},  {\em Nucl.
  Phys.} {\bf {\bf B729}} (2005) 117--134,
  [\href{http://xxx.lanl.gov/abs/hep-lat/0507035}{{\tt hep-lat/0507035}}].

\bibitem{Becirevic:2005ta}
D.~Becirevic {\em et.~al.}, {\it {\it Non-perturbatively renormalised light
  quark masses from a lattice simulation with $N_f = 2$}},  {\em Nucl. Phys.}
  {\bf {\bf B734}} (2006) 138--155,
  [\href{http://xxx.lanl.gov/abs/hep-lat/0510014}{{\tt hep-lat/0510014}}].

\bibitem{Blossier:2007vv}
{\bf European Twisted Mass} Collaboration, B.~Blossier {\em et.~al.}, {\it {\it
  Light quark masses and pseudoscalar decay constants from $N_f=2$ Lattice QCD
  with twisted mass fermions}},  \href{http://xxx.lanl.gov/abs/0709.4574}{{\tt
  arXiv:0709.4574}}.

\bibitem{Aubin:2004ck}
{\bf HPQCD} Collaboration, C.~Aubin {\em et.~al.}, {\it {\it First
  determination of the strange and light quark masses from full lattice QCD}},
  {\em Phys. Rev.} {\bf D {\bf 70}} (2004) 031504,
  [\href{http://xxx.lanl.gov/abs/hep-lat/0405022}{{\tt hep-lat/0405022}}].

\bibitem{Aubin:2004fs}
{\bf MILC} Collaboration, C.~Aubin {\em et.~al.}, {\it {\it Light pseudoscalar
  decay constants, quark masses, and low energy constants from three-flavor
  lattice QCD}},  {\em Phys. Rev.} {\bf D {\bf 70}} (2004) 114501,
  [\href{http://xxx.lanl.gov/abs/hep-lat/0407028}{{\tt hep-lat/0407028}}].

\bibitem{Mason:2005bj}
{\bf HPQCD} Collaboration, Q.~Mason, H.~D. Trottier, R.~Horgan, C.~T.~H.
  Davies, and G.~P. Lepage, {\it {High-precision determination of the
  light-quark masses from realistic lattice QCD}},  {\em Phys. Rev.} {\bf D
  {\bf 73}} (2006) 114501, [\href{http://xxx.lanl.gov/abs/hep-ph/0511160}{{\tt
  hep-ph/0511160}}].

\bibitem{Ishikawa:2007nn}
{\bf JLQCD} Collaboration, T.~Ishikawa {\em et.~al.}, {\it {\it Light quark
  masses from unquenched lattice QCD}},
  \href{http://xxx.lanl.gov/abs/0704.1937}{{\tt arXiv:0704.1937}}.

\bibitem{Duncan:1996xy}
A.~Duncan, E.~Eichten, and H.~Thacker, {\it {\it Electromagnetic Splittings and
  Light Quark Masses in Lattice QCD}},  {\em Phys. Rev. Lett.} {\bf {\bf 76}}
  (1996) 3894--3897, [\href{http://xxx.lanl.gov/abs/hep-lat/9602005}{{\tt
  hep-lat/9602005}}].

\bibitem{Blum:2007cy}
T.~Blum, T.~Doi, M.~Hayakawa, T.~Izubuchi, and N.~Yamada, {\it {\it
  Determination of light quark masses from the electromagnetic splitting of
  pseudoscalar meson masses computed with two flavors of domain wall
  fermions}},  {\em Phys. Rev.} {\bf D {\bf 76}} (2007) 114508,
  [\href{http://xxx.lanl.gov/abs/0708.0484}{{\tt arXiv:0708.0484}}].

\bibitem{Namekawa:2005dr}
Y.~Namekawa and Y.~Kikukawa, {\it {\it Electromagnetic mass difference on the
  lattice}},  {\em PoS} {\bf {\bf LAT2005}} (2006) 090,
  [\href{http://xxx.lanl.gov/abs/hep-lat/0509120}{{\tt hep-lat/0509120}}].

\bibitem{Shintani:2007ub}
{\bf JLQCD} Collaboration, E.~Shintani {\em et.~al.}, {\it {\it Pion mass
  difference from vacuum polarization}},  {\em PoS} {\bf {\bf LAT2007}} (2007)
  134, [\href{http://xxx.lanl.gov/abs/0710.0691}{{\tt arXiv:0710.0691}}].

\bibitem{Das:1967it}
T.~Das, G.~S. Guralnik, V.~S. Mathur, F.~E. Low, and J.~E. Young, {\it {\it
  Electromagnetic mass difference of pions}},  {\em Phys. Rev. Lett.} {\bf {\bf
  18}} (1967) 759--761.

\bibitem{Luscher:1983rk}
M.~Luscher, {\it {\it On a relation between finite size effects and elastic
  scattering processes}}, . Lecture given at Cargese Summer Inst., Cargese,
  France, Sep 1-15, 1983.

\bibitem{Luscher:1985dn}
M.~Luscher, {\it {\it Volume Dependence of the Energy Spectrum in Massive
  Quantum Field Theories. 1. Stable Particle States}},  {\em Commun. Math.
  Phys.} {\bf {\bf 104}} (1986) 177.

\bibitem{Luscher:1986pf}
M.~Luscher, {\it {\it Volume Dependence of the Energy Spectrum in Massive
  Quantum Field Theories. 2. Scattering States}},  {\em Commun. Math. Phys.}
  {\bf {\bf 105}} (1986) 153--188.

\bibitem{Gasser:1987zq}
J.~Gasser and H.~Leutwyler, {\it {\it Spontaneously Broken Symmetries:
  Eeffective Lagrangians at Finite Volume}},  {\em Nucl. Phys.} {\bf {\bf
  B307}} (1988) 763.

\bibitem{Hasenfratz:1989pk}
P.~Hasenfratz and H.~Leutwyler, {\it {\it Goldstone boson related finite size
  effects in field theory and critical phenomena with $O(N)$ symmetry}},  {\em
  Nucl. Phys.} {\bf {\bf B343}} (1990) 241--284.

\bibitem{Hansen:1990un}
F.~C. Hansen, {\it {\it Finite size effects in spontaneously broken $SU(N)
  \times SU(N)$ theories}},  {\em Nucl. Phys.} {\bf {\bf B345}} (1990)
  685--708.

\bibitem{Colangelo:2003hf}
G.~Colangelo and S.~Durr, {\it {\it The pion mass in finite volume}},  {\em
  Eur. Phys. J.} {\bf C {\bf 33}} (2004) 543--553,
  [\href{http://xxx.lanl.gov/abs/hep-lat/0311023}{{\tt hep-lat/0311023}}].

\bibitem{AliKhan:2003cu}
{\bf QCDSF-UKQCD} Collaboration, A.~Ali~Khan {\em et.~al.}, {\it {\it The
  nucleon mass in $N_F = 2$ lattice QCD: Finite size effects from chiral
  perturbation theory}},  {\em Nucl. Phys.} {\bf {\bf B689}} (2004) 175--194,
  [\href{http://xxx.lanl.gov/abs/hep-lat/0312030}{{\tt hep-lat/0312030}}].

\bibitem{Koma:2004wz}
Y.~Koma and M.~Koma, {\it {\it On the finite size mass shift formula for stable
  particles}},  {\em Nucl. Phys.} {\bf {\bf B713}} (2005) 575--597,
  [\href{http://xxx.lanl.gov/abs/hep-lat/0406034}{{\tt hep-lat/0406034}}].

\bibitem{Colangelo:2004sc}
G.~Colangelo, {\it {\it Finite volume effects in chiral perturbation theory}},
  {\em Nucl. Phys. Proc. Suppl.} {\bf {\bf 140}} (2005) 120--126,
  [\href{http://xxx.lanl.gov/abs/hep-lat/0409111}{{\tt hep-lat/0409111}}].

\bibitem{Colangelo:2005gd}
G.~Colangelo, S.~Durr, and C.~Haefeli, {\it {\it Finite volume effects for
  meson masses and decay constants}},  {\em Nucl. Phys.} {\bf {\bf B721}}
  (2005) 136--174, [\href{http://xxx.lanl.gov/abs/hep-lat/0503014}{{\tt
  hep-lat/0503014}}].

\bibitem{Caprini:2005zr}
I.~Caprini, G.~Colangelo, and H.~Leutwyler, {\it {\it Mass and width of the
  lowest resonance in QCD}},  {\em Phys. Rev. Lett.} {\bf {\bf 96}} (2006)
  132001, [\href{http://xxx.lanl.gov/abs/hep-ph/0512364}{{\tt
  hep-ph/0512364}}].

\bibitem{Colangelo:2006mp}
G.~Colangelo and C.~Haefeli, {\it {\it Finite volume effects for the pion mass
  at two loops}},  {\em Nucl. Phys.} {\bf {\bf B744}} (2006) 14--33,
  [\href{http://xxx.lanl.gov/abs/hep-lat/0602017}{{\tt hep-lat/0602017}}].

\bibitem{Bardeen:1988zw}
W.~A. Bardeen, J.~Bijnens, and J.~M. Gerard, {\it {\it Hadronic Matrix Elements
  and the $\pi^+$-$\pi^0$ Mass Difference}},  {\em Phys. Rev. Lett.} {\bf {\bf
  62}} (1989) 1343.

\bibitem{Ecker:1988te}
G.~Ecker, J.~Gasser, A.~Pich, and E.~de~Rafael, {\it {\it The Role of
  Resonances in Chiral Perturbation Theory}},  {\em Nucl. Phys.} {\bf {\bf
  B321}} (1989) 311.

\bibitem{Urech:1994hd}
R.~Urech, {\it {\it Virtual photons in chiral perturbation theory}},  {\em
  Nucl. Phys.} {\bf {\bf B433}} (1995) 234--254,
  [\href{http://xxx.lanl.gov/abs/hep-ph/9405341}{{\tt hep-ph/9405341}}].

\bibitem{Knecht:1997jw}
M.~Knecht and R.~Urech, {\it {\it Virtual photons in low energy $\pi$-$\pi$
  scattering}},  {\em Nucl. Phys.} {\bf {\bf B519}} (1998) 329--360,
  [\href{http://xxx.lanl.gov/abs/hep-ph/9709348}{{\tt hep-ph/9709348}}].

\bibitem{Bijnens:2006mk}
J.~Bijnens and N.~Danielsson, {\it {\it Electromagnetic corrections in
  partially quenched chiral perturbation theory}},  {\em Phys. Rev.} {\bf D
  {\bf 75}} (2007) 014505, [\href{http://xxx.lanl.gov/abs/hep-lat/0610127}{{\tt
  hep-lat/0610127}}].

\bibitem{Haefeli:2007ey}
C.~Haefeli, M.~A. Ivanov, and M.~Schmid, {\it {\it Electromagnetic low-energy
  constants in ChPT}},  {\em Eur. Phys. J.} {\bf C {\bf 53}} (2008) 549--557,
  [\href{http://xxx.lanl.gov/abs/0710.5432}{{\tt 0710.5432}}].

\bibitem{Gasser:2003hk}
J.~Gasser, A.~Rusetsky, and I.~Scimemi, {\it {\it Electromagnetic corrections
  in hadronic processes}},  {\em Eur. Phys. J.} {\bf {\bf C32}} (2003) 97--114,
  [\href{http://xxx.lanl.gov/abs/hep-ph/0305260}{{\tt hep-ph/0305260}}].

\bibitem{Bernard:1992mk}
C.~W. Bernard and M.~F.~L. Golterman, {\it {\it Chiral perturbation theory for
  the quenched approximation of QCD}},  {\em Phys. Rev.} {\bf D {\bf 46}}
  (1992) 853--857, [\href{http://xxx.lanl.gov/abs/hep-lat/9204007}{{\tt
  hep-lat/9204007}}].

\bibitem{Sharpe:1992ft}
S.~R. Sharpe, {\it {\it Quenched chiral logarithms}},  {\em Phys. Rev.} {\bf D
  {\bf 46}} (1992) 3146--3168,
  [\href{http://xxx.lanl.gov/abs/hep-lat/9205020}{{\tt hep-lat/9205020}}].

\bibitem{Sharpe:2000bc}
S.~R. Sharpe and N.~Shoresh, {\it {\it Physical results from unphysical
  simulations}},  {\em Phys. Rev.} {\bf D {\bf 62}} (2000) 094503,
  [\href{http://xxx.lanl.gov/abs/hep-lat/0006017}{{\tt hep-lat/0006017}}].

\bibitem{Sharpe:2001fh}
S.~R. Sharpe and N.~Shoresh, {\it {\it Partially quenched chiral perturbation
  theory without $\Phi_0$}},  {\em Phys. Rev.} {\bf D {\bf 64}} (2001) 114510,
  [\href{http://xxx.lanl.gov/abs/hep-lat/0108003}{{\tt hep-lat/0108003}}].

\bibitem{Henneaux:1992ig}
M.~Henneaux and C.~Teitelboim, {\it {\it Quantization of gauge systems}}, .
  Princeton University Presse, New Jersey (1992), 520 p.

\bibitem{Filk:1996dm}
T.~Filk, {\it {\it Divergencies in a field theory on quantum space}},  {\em
  Phys. Lett.} {\bf {\bf B376}} (1996) 53--58.

\bibitem{Minwalla:1999px}
S.~Minwalla, M.~Van~Raamsdonk, and N.~Seiberg, {\it {\it Noncommutative
  perturbative dynamics}},  {\em JHEP} {\bf 02} (2000) 020,
  [\href{http://xxx.lanl.gov/abs/hep-th/9912072}{{\tt hep-th/9912072}}].

\bibitem{Hayakawa:1999yt}
M.~Hayakawa, {\it {\it Perturbative analysis on infrared aspects of
  noncommutative QED on $\mathbb{R}^4$}},  {\em Phys. Lett.} {\bf {\bf B478}}
  (2000) 394--400, [\href{http://xxx.lanl.gov/abs/hep-th/9912094}{{\tt
  hep-th/9912094}}].

\bibitem{Bijnens:2006zp}
J.~Bijnens, {\it {\it Chiral perturbation theory beyond one loop}},  {\em Prog.
  Part. Nucl. Phys.} {\bf {\bf 58}} (2007) 521--586,
  [\href{http://xxx.lanl.gov/abs/hep-ph/0604043}{{\tt hep-ph/0604043}}].

\bibitem{Gasser:1983yg}
J.~Gasser and H.~Leutwyler, {\it {\it Chiral Perturbation Theory to One Loop}},
   {\em Ann. Phys.} {\bf {\bf 158}} (1984) 142.

\bibitem{Gasser:1984gg}
J.~Gasser and H.~Leutwyler, {\it {\it Chiral Perturbation Theory: Expansions in
  the Mass of the Strange Quark}},  {\em Nucl. Phys.} {\bf {\bf B250}} (1985)
  465.

\bibitem{Bijnens:1999hw}
J.~Bijnens, G.~Colangelo, and G.~Ecker, {\it {\it Renormalization of chiral
  perturbation theory to order $p^6$}},  {\em Annals Phys.} {\bf {\bf 280}}
  (2000) 100--139, [\href{http://xxx.lanl.gov/abs/hep-ph/9907333}{{\tt
  hep-ph/9907333}}].

\bibitem{Bijnens:1996kk}
J.~Bijnens and J.~Prades, {\it {\it Electromagnetic corrections for pions and
  kaons: Masses and polarizabilities}},  {\em Nucl. Phys.} {\bf {\bf B490}}
  (1997) 239--271, [\href{http://xxx.lanl.gov/abs/hep-ph/9610360}{{\tt
  hep-ph/9610360}}].

\bibitem{Hayakawa:1999zf}
M.~Hayakawa, {\it {\it Perturbative analysis on infrared and ultraviolet
  aspects of noncommutative QED on $\mathbb{R}^4$}},
  \href{http://xxx.lanl.gov/abs/hep-th/9912167}{{\tt hep-th/9912167}}.

\end{thebibliography}\endgroup


\end{document}